\documentclass[aps,prd,eqsecnum,reprint,amsmath,amssymb,amsfonts,nofootinbib,superscriptaddress]{revtex4-2}

\usepackage{graphicx, hyperref, tabularx}
\usepackage{microtype}

\usepackage{bbold}
\renewcommand{\t}{{\mathbb{t}}}  
\newcommand{\one}{{\mathbb{1}}}  
\newcommand{\Rbf}{\mbox{$\mathbf{R}$}}

\newcommand{\xbf}{\mbox{$\mathbf{x}$}}
\newcommand{\lbf}{\mbox{$\mathbf{l}$}}
\newcommand{\Lbf}{\mbox{$\mathbf{L}$}}
\newcommand{\pd}[2]{\dfrac{\partial #1}{\partial #2}}

\DeclareMathOperator{\rank}{rank}

\usepackage{color, xcolor}
\definecolor{mygray}{gray}{0.9}

\newcommand{\apjs}{Astrophys. J. Suppl.}
\newcommand{\lrr}{Living Rev. Rel.}

\usepackage{orcidlink}

\begin{document}



\title{Characteristic Decomposition for Relativistic Numerical Simulations:\\
I. Hydrodynamics}

\author{Saul A. Teukolsky
\orcidlink{0000-0001-9765-4526}}
\email{saul@astro.cornell.edu}
\affiliation{Cornell Center for Astrophysics
   and Planetary Science, Cornell University, Ithaca, NY 14853, USA}
\affiliation{Theoretical Astrophysics 350-17,
  California Institute of Technology, Pasadena, CA 91125, USA}
\date{\today}

\begin{abstract}
The characteristic decomposition for GRMHD is not
known in a form useful for current numerical simulations.
This prevents us from using the most accurate known computational
methods, such as full-wave Riemann solvers. 
In this paper, we present a new method of finding
decompositions. The method is based on transformations
from the comoving frame, where the fluid flow is simplest and
the decomposition has been known for a long time.
The key innovation we introduce is that of
quasi-invertible transformations.
In this first paper, we introduce these transformations
using the simpler example of relativistic hydrodynamics.
We recover the known decomposition for relativistic hydrodynamics
in somewhat simpler form than previously derived, and without the need for
computer algebra.
A new result in this paper is the characteristic
decomposition when the the evolution tracks the composition of
a fluid in nuclear statistical equilibrium.
In Paper II of this series, we apply a quasi-invertible transformation
to derive the complete characteristic decomposition for GRMHD in the
conserved variables used in simulations.
\end{abstract}

\maketitle

\section{Introduction}

Modern computer codes for solving large-scale problems in areas like
fluid dynamics or magnetohydrodynamics (MHD) deal with hyperbolic
evolution equations. Hyperbolicity means that at every point the
flow admits a decomposition into independent waves, each with its
own wave speed.
Formally, the decomposition is carried out by solving an
eigenvalue problem, as described below.
The most accurate numerical algorithms typically
make use in some way of this so-called characteristic decomposition.
For example, boundary conditions can easily and rigorously be imposed
by giving conditions on characteristic variables coming into a domain
and not on those leaving the domain. Similarly, shock waves can
be treated accurately by using relations between the characteristic waves
at the shock discontinuity. Relatedly, the most robust and accurate
prescriptions for numerical fluxes across subdomain boundaries
are based on characteristic decomposition.
While the decomposition is known for relativistic hydrodynamics,
this is not the case for GRMHD for the equations that are currently
used in numerical simulations.

Finding the characteristic decomposition for hydrodynamics or MHD in
full general relativity is considerably more difficult than in Newtonian
physics. First, there is the choice of reference frame.
We want the characteristic decomposition in the coordinates used
in the computer evolution, coordinates
that are often unrelated to the flow itself.
Examples include 
the coordinates of some background
gravitational field such as the Kerr metric,
or a coordinate choice in which Einstein's equations
for the gravitational field can be evolved stably.
The characteristic
decomposition, meanwhile, is simplest in a frame comoving with the
fluid (Lagrangian formulation). And by analogy with Newtonian codes,
we might also want to consider the equivalent of an Eulerian frame.
We will give precise definitions of coordinate, Eulerian, and Lagrangian
observers in relativity in \S\ref{sec:observers}. We will see that relativistic
effects make the relation between these frames much more complicated
than in Newtonian physics.

The second reason for the difficulty of the characteristic
decomposition problem is that the
relativistic equations themselves are more complicated than their
Newtonian counterparts.
The most powerful general numerical algorithms
for problems that admit shocks and surfaces involve writing the
equations in conservation form, where the spatial derivatives of the
evolved quantities enter only as the divergences of fluxes.
So we want the characteristic decomposition in terms of these conserved
variables, not some set of ``primitive'' variables used
in a generic formulation of the equations. The relation between
primitive and conserved variables is again much more complicated
in relativity than in Newtonian physics, especially for MHD.

A straightforward way to proceed is to write down the equations
for the conserved variables in an arbitrary coordinate frame and
find the characteristic decomposition directly. This is actually feasible
for hydrodynamics, especially with the aid of computer algebra.
However, the results do not typically emerge in the simplest form.
Moreover, this method seems infeasible for GRMHD.

More promising is to use a transformation method: One starts with a
tractable formulation, say using a set of primitive variables in the
comoving frame. Then one transforms the decomposition
first to the coordinate frame and then to the conserved variables.
A number of previous authors have tackled hydrodynamics and GRMHD
in this way, as we will describe.
However, the complete decomposition for GRMHD is still not known
because the existing transformation techniques are not powerful
enough.

This paper is the first of two in which
we introduce a new transformation technique, which
we call quasi-invertible transformations. In this first
paper, we describe the technique
by determining the characteristic decomposition for relativistic
hydrodynamics. We recover the known results in somewhat simpler form
and without the need for computer algebra.
Then, in Paper II,
we apply the new technique to find the decomposition
for GRMHD. These results mean that shocks and other discontinuities in
MHD systems can now be handled with the most accurate known
computational methods,
and that we no longer have to rely on approximate treatments of
numerical fluxes and boundary conditions in GRMHD.

A new result in this paper is the characteristic decomposition
when the the evolution tracks the composition
of a fluid in nuclear statistical equilibrium. The transformation
methods we use allow these results to be obtained from simple
extensions of the results where composition changes are ignored.

To set the current paper
in context, we describe previous work on the characteristic
decomposition for the conserved variables
for relativistic hydrodynamics. The complete decomposition
for special relativity was given by Donat et al.~\cite{donat1998}.
It turns out that this treatment can be turned into a general
relativistic version by writing the vector components as spatial
3-vectors in the Eulerian frame.
The first fully 3-dimensional
treatment explicitly in general relativity for the conserved variables was
by Banyuls et al.~\cite{banyuls1997}. They gave only the right eigenvectors
of the decomposition along the $x$-axis, with the decomposition along the
other spatial axes being given by suitable permutations. These results
were extended by Ib\'a\~nez et al.~\cite{ibanez2001} to include the left
eigenvectors. The decomposition was reproduced in the comprehensive
review article~\cite{font2008}, with a simpler version presented in the
updated review~\cite{marti2015}\footnote{This version uses a different 
definition
of the conserved variable $\tau$ than the earlier treatments, resulting in
some small differences.}. A more complete review of related early work,
including developments in special relativity, is given in Ref.~\cite{font2000}.

There are two other efforts that are related to the work presented here.
The first is the treatment of the characteristic decomposition for
MHD in special relativity given in Refs.\ \cite{antonthesis,anton2010},
which also uses a transformation method. 
In an appendix to Paper II, we describe the drawbacks of this method and
how they are overcome with quasi-invertible transformations.
The second
recent work is that of Refs.\ \cite{schoepe2018,hilditch2019}.
While these authors focused on MHD, this time in general relativity,
they also applied their transformation method to hydrodynamics.
We also discuss
the relation to our methods in an appendix in Paper II.

This paper is structured as follows:
In \S\ref{sec:decomp} we describe the characteristic decomposition problem
and the covariant form of the associated eigenvector problem.
\S\ref{3plus1} uses the 3+1 decomposition to relate the coordinate,
Eulerian, and Lagrangian formulations of evolution equations. 
\S\ref{sec:comoving} derives the characteristic decomposition for
hydrodynamics in the Lagrangian (comoving) description.
In \S\ref{sec:changing} we introduce the simplest case of changing
the variables used in a formulation.
\S\ref{sec:eulerian} discusses the important case of transforming from
4-velocity to 3-velocity. Since the Jacobian matrix
of the transformation is $4\times 3$, it is not strictly speaking
invertible. We discuss how to define a useful ``inverse'' transformation
matrix that plays an important role in the subsequent developments.
\S\ref{sec:lefttrans} introduces the key idea of this paper, the
concept of a quasi-invertible transformation, of which the transformation
between 4-velocity and 3-velocity is a simple example.
\S\ref{sec:eulerianframe} applies the quasi-invertible transformation for
velocities to transform the comoving decomposition of
\S\ref{sec:comoving} to the Eulerian decomposition.
Then \S\ref{sec:conservative} transforms the Eulerian decomposition to
conservative variables, which is what is needed for numerical codes.
\S\ref{sec:composition} extends the characteristic decomposition to
include composition dependence for the fluid, and finally
\S\ref{sec:discussion} discusses and summarizes the paper.

\section{The Characteristic Decomposition Problem}
\label{sec:decomp}

In this work we will be interested in quasi-linear systems of the form
\begin{equation}
\frac{\partial U^\alpha}{\partial t}+
A^{k\alpha}{}_\beta\frac{\partial U^\beta}{\partial x^k}=S^\alpha.
\label{eq:qlinear}
\end{equation}
Here $U^\alpha$ is the vector of dynamical fields. The matrices
$A^{k\alpha}{}_\beta$ and the source vector $S^\alpha$ may depend on
$U^\alpha$ but not its derivatives. Greek letters $\alpha,\beta,\ldots$
label the collection of dynamical fields. Roman letters $a,b,\ldots$
from the beginning of the alphabet label 4-d abstract spacetime indices.
Roman letters $i,j,\ldots$ from the middle of the alphabet
label 3-d spatial indices in some particular coordinate system.
When there is no possibility of confusion, we will omit the indices
$\alpha,\beta,\ldots$, and matrix or matrix-vector multiplication
over the system quantities will be understood.

Note that since we have a metric in general relativity, there is
no practical difference in denoting a vector as $f^a$ or $f_a$.
With abstract indices, both quantities denote a rank-1 tensor as a geometric
object, not its components in some basis.

The characteristic decomposition of Eq.\ \eqref{eq:qlinear} depends
only on the principal part of the equation (i.e., the part containing
derivatives), so we can ignore $S^\alpha$. We will ignore similar source
terms throughout this paper
as we determine various characteristic decompositions.

For a problem specified in conservation form, the evolution equations
take the form
\begin{equation}
\frac{\partial U^\alpha}{\partial t} + \frac{\partial}{\partial x^k}
\left[F^{k\alpha}(U^\beta)\right]=S^\alpha,
\end{equation}
where $F^{k\alpha}$ is the $k$-component of the flux for the $\alpha$
variable. We see that this equation
can be written in the form  \eqref{eq:qlinear} with the characteristic
matrix equal to the Jacobian matrix of the flux:
\begin{equation}
A^{k\alpha}{}_\beta=\frac{\partial F^{k\alpha}}{\partial U^\beta}.
\end{equation}

A characteristic decomposition is defined with respect to
a specific spatial direction $\zeta^k$. Typically, this is the outward unit
normal vector to a boundary. If we consider Eq.\ \eqref{eq:qlinear}
locally, so that the matrices $A^k$ can be treated as constant, then
there are solutions of the form
\begin{equation}
X_t(t,x^k)=X\exp[i(\zeta_k x^k-y t)].
\end{equation}
Substituting this ansatz into Eq.\ \eqref{eq:qlinear}, we see that the
characteristic speeds are the eigenvalues $y$ of the equation
\begin{equation}
-y X + (\zeta_k A^k)X =0.
\label{eq:right}
\end{equation}
Here $X$ is a right eigenvector (components $X^\alpha$).
The left eigenvectors satisfy
\begin{equation}
-y L + L(\zeta_k A^k) =0.
\label{eq:left}
\end{equation}
Strongly hyperbolic systems, which we are considering here, by definition
have a complete eigensystem, so the matrix $[X]$ of $X$'s (as columns)
is invertible with inverse equal to $[L]$, the matrix of $L$'s (as rows).
The transformations to characteristic variables and back again are given by
\begin{equation}
U_{\rm char}=[L]U,\quad U=[X] U_{\rm char}
\end{equation}
and play an important role in many numerical algorithms for solving
equations of the form \eqref{eq:qlinear}.

Sometimes we encounter equations where the coefficient of the
$\partial/\partial t$ term in Eq.\ \eqref{eq:qlinear}
is not the identity matrix:
\begin{equation}
A^0\frac{\partial U}{\partial t}+
A^k\frac{\partial U}{\partial x^k}=S.
\label{eq:general_qlinear}
\end{equation}
Such equations lead to a generalized eigenvalue problem, where the
coefficient of $y$ in Eq.\ \eqref{eq:right} or \eqref{eq:left} is multiplied
by $A^0$. Equivalently, we get a standard eigenvalue problem by multiplying
Eq.\ \eqref{eq:general_qlinear}
through by $(A^0)^{-1}$.

We can write Eq.\ \eqref{eq:qlinear} or \eqref{eq:general_qlinear}
covariantly as
\begin{equation}
A^a\nabla_a U=0.
\label{eq:system}
\end{equation}
The principal part of this equation is the same as that of the
coordinate-specific forms given above, since we can regard
terms containing connection coefficients as source terms. We now follow
Refs.\ \cite{Friedrichs1974,anile1989}
and introduce a unit timelike vector $\xi_a$
along with the unit spacelike vector $\zeta_a$. Then, if we define
\begin{equation}
q_a=y \xi_a+\zeta_a,
\label{qalpha}
\end{equation}
the eigenvalue problem for Eq.\ \eqref{eq:system}
for the direction $\zeta_a$ can be written as
\begin{equation}
(A^a q_a)X=0,
\label{eq:eigenvalue}
\end{equation}
where $y$ is the eigenvalue and $X$ is the corresponding eigenvector.
To find the eigenvalues from Eq.\ \eqref{eq:eigenvalue},
we must solve the equation
\begin{equation}
\det(A^a q_a) = 0.
\label{eq:det_eqns}
\end{equation}

Note that any reformulation of the system of equations derived by making
linear combinations of the original equations will have the same
characteristic decomposition. This is because the eigenvectors
of Eq.\ \eqref{eq:eigenvalue} and the determinant in Eq.\ \eqref{eq:det_eqns}
are invariant under forming linear combinations of the rows of the matrix
$A^a q_a$.

In Minkowski spacetime in Cartesian coordinates,
if we choose $\xi^a=(1,0,0,0)$ and $\zeta^a=(0,1,0,0)$,
Eq.\ \eqref{eq:eigenvalue} gives the usual (generalized) eigenvalue problem
\begin{equation}
A^x X = y A^0 X.
\label{eq:generalized}
\end{equation}
For an arbitrary spacetime, without
loss of generality we can choose $\zeta^a$ such that
$\zeta^a \xi_a=0$ (any nonzero component of $\zeta^a$
along $\xi^a$ can be subtracted and then $\zeta^a$ can
be renormalized).
Two specific choices of $\xi^a$ will be relevant in this paper.
First, we will take $\xi^a=u^a$, the 4-velocity of the fluid,
and describe the eigenvalue problem
for comoving observers. In this case, we will denote the
corresponding eigenvalue in Eq.\ \eqref{qalpha} by $y_u$.
Later, we will take $\xi^a=n^a$, the normal
to some $t=\text{constant}$ slice, and
discuss Eulerian observers, the case most relevant for numerical codes.
In this case, we will write the corresponding eigenvalue simply as $y$.

Note that there is an important case in which the characteristic
decomposition for hydrodynamics or MHD decouples from the decomposition
for Einstein's equations. This occurs when Einstein's equations are
reduced to first-order form with the stress-energy tensor $T^{ab}$ a
source term. The tensor $T^{ab}$ often does not contain derivatives
of the matter variables. Similarly, the matter equations typically
do not contain second derivatives of the metric, that is, first
derivatives of the reduction variables introduced
in Einstein's equations. Accordingly, the principal parts of
the characteristic matrices
for gravity and matter are decoupled and can be treated separately.
For further discussion of this point, see \S II.B.3 of Ref.~\cite{schoepe2018}.

\section{\boldmath{$3+1$} Decomposition}
\label{3plus1}

\subsection{Coordinate, Eulerian, and Lagrangian Observers}
\label{sec:observers}
In a relativistic
 evolution code, we choose a time slicing (a set of $t=\text{constant}$
slices) and regard the evolution as taking place on the slices, one after
another, advancing into the future. Equivalently, we can consider
a non-normal congruence
of curves threading the slices, with each curve labeled by its
spatial coordinates on the initial slice. We then interpret these curves
as worldlines of
a set of ``coordinate observers''
with tangent vectors $t^a=(\partial/\partial t)^a$.

As a special case, we can consider a set of coordinate observers
each with 4-velocity $n^a$ orthogonal
to the time slices.
The evolution equations would then describe the physics
in the reference frame of these observers, who are at rest in the time
slices. By analogy with non-relativistic hydrodynamics, these
observers are called Eulerian observers (cf.\ \cite{smarryork1978}).
Coordinate observers (tangent vector $t^a$)
are related to Eulerian observers (normalized tangent vector $n^a$)
by a velocity shift at each
point on the slice.
The relationship is specified by
\begin{equation}
t^a = \alpha n^a+\beta^a.
\label{eq:tdecomp}
\end{equation}
Here $\alpha$ is the lapse function and $\beta^a$ is the shift vector,
which is spatial (orthogonal to $n^a$).

As another special case,
if we choose the shift so that the spatial coordinates along
the worldlines of a fluid are constant,
then $t^a$ would be proportional
to $u^a$, the 4-velocity of the fluid.
These observers would
be Lagrangian observers since they are comoving with the fluid.

\subsection{\boldmath{$3+1$} Quantities}
In the
$3+1$ coordinate system adapted to the slicing defined by
$t$ = constant, the metric tensor is
\begin{equation}
ds^2=g_{ab}dx^a dx^b
=-\alpha^2 dt^2+\gamma_{ij}(dx^i+\beta^i dt)(dx^j +\beta^j dt),
\label{eq:3+1}
\end{equation}
or
\begin{equation}
\begin{split}
g_{ab}&=\begin{pmatrix}
-\alpha^2 +\beta_i\beta^i & \beta_i\\
\beta_j & \gamma_{ij}
\end{pmatrix},\\
g^{ab}&=\begin{pmatrix}
-\alpha^{-2} & \alpha^{-2}\beta^i\\
\alpha^{-2}\beta^j & \gamma^{ij}-\alpha^{-2}\beta^i\beta^j
\end{pmatrix}.
\end{split}
\label{eq:metric}
\end{equation}
Here
$\gamma_{ij}$ is the spatial metric on the $t=\text{constant}$ slices.
The unit normal vector $n^a$ has components
\begin{equation}
n_a=(-\alpha,0,0,0),\quad n^a=(1/\alpha,-\beta^i/\alpha).
\label{eq:normal}
\end{equation}

For a more complete treatment of the $3+1$ decomposition,
see, e.g., \cite{baumgarteShapiroBook,rezzollabook}.

\subsection{Relation Between Coordinate and Eulerian Formulations}

In general, the equations in numerical evolution codes
are written in some coordinate system with the time derivative operator
\begin{equation}
\frac{\partial}{\partial t}=\mathcal{L}_{t^a}+\text{non-derivative terms,}
\end{equation}
where $\mathcal{L}_{t^a}$ denotes a Lie derivative.
By Eq.\ \eqref{eq:tdecomp} we have
\begin{equation}
\mathcal{L}_{t^a}=\alpha\mathcal{L}_{n^a}+\mathcal{L}_{\beta^a}
+\text{$\alpha$-derivatives.}
\label{eq:liederiv}
\end{equation}
Since the shift term is purely a spatial derivative,
we see that if we can find the characteristic decomposition for the
evolution operator $\mathcal{L}_{n^a}$, then the decomposition for the operator
$\mathcal{L}_{t^a}$ has the same eigenvectors but the eigenvalues are
related by
\begin{equation}
y_t=\alpha y-\beta_n.
\label{eq:eigs_related}
\end{equation}
Here $\beta_n$ is the projection of $\beta^a$ along the spatial normal vector
$s^a$ for which the decomposition is being performed and $y$ is the eigenvalue
in the Eulerian frame.
So we will focus on determining the characteristic decomposition
with respect to $n^a$, that is, in the
rest frame of the Eulerian  observer.

\subsection{Eulerian and Lagrangian formulations}

In Newtonian hydrodynamics, the Eulerian formulation uses the partial
derivative $\partial/\partial t$ for the time evolution.  The Lagrangian
formulation uses the comoving time derivative $d/dt=\partial/\partial t
+v^i\partial/\partial x^i$. Nevertheless, the characteristic
decompositions are simply related: the Eulerian characteristic speeds
are equal to the Lagrangian speeds plus the normal component of the
fluid velocity. The eigenvectors are the same in both cases.
This relationship is mathematically similar to the effect of the shift
vector in Eqs.~\eqref{eq:liederiv} and \eqref{eq:eigs_related}.

In the relativistic case, the decompositions are not the same.
The time derivative for the Lagrangian formulation
is $u^a \nabla_a\sim d/d\tau$,
where $\tau$ is the proper time of the observer comoving with the
fluid.\footnote{%
We use the tilde because the relation is an equality only for scalars
or for special relativistic vectors. For vectors in general relativity,
or in curvilinear coordinates, there are additional terms from the
connection coefficients. However, these do not affect the characteristic
decomposition since they do not involve derivatives of the variables.}
The relation between $d/d\tau$ and $\partial/\partial t$ is not
a simple one analogous to
the relation
\eqref{eq:liederiv} between the coordinate and Eulerian formulations,
and is partly responsible for the complexity of the relativistic characteristic
decomposition.

\section{Comoving Decomposition}
\label{sec:comoving}

\subsection{Primitive and Conserved Variables}

As one might expect, the characteristic decomposition is most
easily determined in the comoving (Lagrangian) frame, since
that is the natural frame associated with the fluid flow.
The formulation of the fundamental equations in this frame
is not in conservation form (i.e., derivatives of the variables
do not appear only as divergences of fluxes).
To handle shocks satisfactorily, most current numerical
codes use a formulation of the fundamental equations that
is in conservation form, and the evolved variables are called
conserved variables. As mentioned earlier, finding the characteristic
decomposition directly for such a formulation is complicated.
Instead, it turns out to be simpler to first find the
decomposition for a set of primitive variables and
then transform the eigenvectors to
the conserved variables. (The result is
independent of the original choice of primitive variables.)
The simplest starting approach is to use $u^a$ as one of the
primitive variables. In addition, assuming we are dealing with a
one-component perfect fluid with constant composition,
we need two variables describing the
thermodynamic state of the fluid.
Of course, at some stage we have to enforce the constraint $u^a u_a=-1$,
reducing the number of independent variables from 6 to 5.

The treatment in this section is based on that of Anile~\cite{anile1989},
except that we 
work with 5-d eigenvectors and avoid the spurious eigenvector
present in Anile's 6-component treatment.

It does not make much difference which two thermodynamic variables
we choose as primitive variables; it is easy to transform from
one set to any other, as we will see. We will choose the pair
$(p,\epsilon)$, where $p$ is the pressure and $\epsilon$ the
specific internal energy density. This choice simplifies the
algebra slightly.
Later, we will replace $p$ by $\rho$, the rest-mass density,
since $(\rho,\epsilon)$ are the variables commonly used
in astrophysical codes.

Associated with the fluid are also
the total energy density $e$,
the specific entropy $s$, 
and the specific enthalpy $h$.
These quantities are related by the following
equations:
\begin{align}
\label{eq:e}
e&=\rho(1+\epsilon),\\
h&=1+\epsilon+p/\rho,\\
\label{eq:ds}
T\,ds&=d\epsilon+p\,d(1/\rho)\quad\text{(First law),}\\
p&=p(\rho,\epsilon)\qquad\text{(Equation of state),}\\
dp&=
\left.\frac{\partial p}{\partial \rho}\right|_\epsilon d\rho +
\left.\frac{\partial p}{\partial \epsilon}\right|_\rho d\epsilon 
\equiv \chi d\rho + \kappa d\epsilon.
\label{eq:chi}
\end{align}
Here $T$ is the temperature.
As shown in Appendix \ref{app:a}, the quantities $\chi$ and
$\kappa$ are related
to the speed of sound $c_s$ by
\begin{equation}
c_s^2=\frac{1}{h}\left(\chi+\frac{p}{\rho^2}\kappa\right).
\label{eq:cs}
\end{equation}

With these variables, the equations of hydrodynamics are the
Euler equation
\begin{equation}
\rho h\, u^a\nabla_a u^b+h^{ab}\nabla_a p=0,
\label{eq:euler}
\end{equation}
and the equation of mass conservation,
\begin{equation}
u^a\nabla_a \rho+\rho \nabla_a u^a=0.
\label{eq:rhocons}
\end{equation}
Here $h^{ab}$ is the projection tensor that projects orthogonal
to $u^a$:
\begin{equation}
h^{ab}=g^{ab}+u^a u^b.
\end{equation}
Equivalently, $h^{ab}$ is the 3-dimensional metric tensor in the comoving
frame, analogous to $\gamma^{ab}$ for the Eulerian frame.

The final equation is the energy equation. Since the fluid is
taken to be perfect, the flow is isentropic (except at shocks).
Accordingly, we set $ds=0$ in Eq.\ \eqref{eq:ds} and get
\begin{equation}
\frac{d\epsilon}{d\tau}=\frac{p}{\rho^2}\frac{d\rho}{d\tau}.
\label{eq:epsrho}
\end{equation}
Here $\tau$ is the proper time along the fluid worldlines. Rewrite
$d/d\tau=u^a \nabla_a$ and use Eq.\ \eqref{eq:rhocons}
so that the energy equation becomes
\begin{equation}
u^a\nabla_a \epsilon+\frac{p}{\rho}\nabla_a u^a=0.
\label{eq:epsilon}
\end{equation}

To use $p$ as a primitive variable, we need its evolution equation.
Equation \eqref{eq:chi} gives
\begin{align}
\frac{dp}{d\tau}&=
\chi \frac{d\rho}{d\tau} + \kappa \frac{d\epsilon}{d\tau}\notag\\
&=\left(\chi+\kappa\frac{p}{\rho^2}\right)\frac{d\rho}{d\tau}\notag\\
&=h c_s^2 \frac{d\rho}{d\tau},
\end{align}
where we have used Eqs.\eqref{eq:epsrho} and \eqref{eq:cs}. So finally
\begin{equation}
u^a\nabla_a p + \rho h c_s^2\,\nabla_a u^a=0.
\label{eq:pressure}
\end{equation}

To enforce the condition $u^a u_a=-1$, note that in the above equations
we can use the identity
\begin{equation}
\nabla_a u^a=h^a{}_b \nabla_a u^b.
\label{eq:div}
\end{equation}
Then Eqs.\ \eqref{eq:euler}, \eqref{eq:pressure} and  \eqref{eq:epsilon}
can be written as a quasi-linear system \eqref{eq:system}
for the variables $(u^a,p,\epsilon)$,
where the matrices $A^a$ are
\begin{equation}
A^a=
\begin{bmatrix}
u^a h_c{}^b & h_c{}^a/(\rho h) & 0\\[3pt]
\rho h c_s^2 h^{ba}     & u^a    & 0\\[3pt]
(p/\rho)h^{ba} & 0    & u^a
\end{bmatrix}.
\label{eq:amatrix}
\end{equation}

\subsection{Characteristic speeds}
\label{sec:charac}
To find the eigenvalues from Eq.\ \eqref{eq:eigenvalue}, we must set
to zero the determinant
\begin{multline}
\det(A^a q_a)=\det
\begin{bmatrix}
u^a q_a h_c{}^b & h_c{}^a q_a/(\rho h) & 0\\[3pt]
\rho h c_s^2 h^{ba}q_a     & u^aq_a    & 0\\[3pt]
(p/\rho)h^{ba}q_a & 0    & u^a q_a
\end{bmatrix}\\[3pt]
\label{eq:line2}
=(u^a q_a)^2 \det\left[
u_a q^a h_c{}^b-c_s^2h_{ca}q^a h^b{}_a q^a/(q_a u^a)\right].
\end{multline}
Equation \eqref{eq:line2} follows from the previous line using
the formula for the determinant of a partitioned matrix:
\begin{equation}
\det
\begin{pmatrix}
P & Q\\
R & S
\end{pmatrix}
=
\det S \det(P-QS^{-1}R),
\label{eq:partition}
\end{equation}
with $P=u^a q_a h_c{}^b$.
The remaining determinant in Eq.\ \eqref{eq:line2} requires some care.
As a 4-d object, the tensor whose determinant we want
only has rank 3 because it is orthogonal to $u^a$.
Thus naively the determinant
is zero. To evaluate the determinant correctly, note
that it is a scalar and so evaluate it
in the local rest frame of the fluid. The projection tensor
$h_c{}^b$ becomes the spatial metric $\delta_i{}^j$, and so we get
\begin{align}
\det\big[
u_a q^a \delta_i{}^j&-c_s^2q_i q^j/(q_a u^a)\big]\notag\\
&= (q_a u^a)^2[(q_a u^a)-c_s^2 q_i q^i/(q_a u^a)]\notag\\
&= (q_a u^a)[(q_a u^a)^2-c_s^2 h_{ab}q^a q^b],
\label{eq:line3}
\end{align}
where in the last line we have gone back to a general 4-d reference frame.
The second line in Eq.~\eqref{eq:line3} follows from the matrix determinant
lemma:
\begin{equation}
\det(z\delta^i{}_j-aw^i v_j)= z^{d-1}(z-a w^i v_i),
\label{eq:determ}
\end{equation}
where $w^i$ and $v^i$ are $d$-dimensional vectors and $z$ and $a$ are scalars.
So finally we get
\begin{equation}
\det(A^a q_a)=
(q^a u_a)^3
  \left[(q^au_a)^2-c_s^2h_{ab}q^a q^b\right].
\label{eq:eigs}
\end{equation}
We see that there are 3 degenerate eigenvalues corresponding to
$q^au_a=0$ and two non-degenerate eigenvalues corresponding to the
roots of the quadratic expression in the square brackets.

Ref.~\cite{anile1989} shows that the eigenvalue $y$ of Eq.~\eqref{qalpha}
is the
3-velocity in the normal direction for a characteristic
hypersurface of the system \eqref{eq:system} as measured
by an observer with 4-velocity $\xi_a$. For the comoving observer
($\xi_a=u_a$), 
\begin{equation}
q^a u_a=(y_u u^a+\zeta^a)u_a=-y_u,
\label{eq:uq}
\end{equation}
which has the value zero for the degenerate eigenvalues.

From Eq.\ \eqref{eq:eigs}, the non-degenerate eigenvalues satisfy
\begin{equation}
  (q^au_a)^2-c_s^2h_{ab}q^a q^b=0.
\label{eq:quadratic}
\end{equation}
This is a quadratic equation in the eigenvalue $y_u$, with two distinct
roots $y_{u\pm}$.
Since
\begin{equation}
h_{ab}q^a q^b=h_{ab}(y_u u^a+\zeta^a)(y_u u^b+\zeta^b)=\zeta^a\zeta_a=1,
\label{eq:qqh}
\end{equation}
the eigenvalue equation \eqref{eq:quadratic} gives
\begin{equation}
y_{u\pm} = \pm c_s.
\label{eq:mucs}
\end{equation}
As in the non-relativistic case, we find that the characteristic
speeds measured by the comoving observer are zero and $\pm c_s$.

\subsection{Right Eigenvectors for the Degenerate Eigenvalues}
From Eq.\ \eqref{eq:eigs} we see that there is a 3-fold degenerate
eigenvalue given by $q^a u_a=0$.
The corresponding right eigenvectors follow from  solving
\begin{align}
0 &= (A^a q_a)X\nonumber\\
&=\begin{bmatrix}
0 & h_c{}^a q_a/(\rho h) & 0\\[3pt]
\rho h c_s^2 h^{ba}q_a     & 0   & 0\\[3pt]
(p/\rho)h^{ba} q_a & 0    & 0
\end{bmatrix}
\begin{bmatrix}
X_b\\
X_4\\
X_5
\end{bmatrix}\nonumber\\
&=\begin{bmatrix}
 h_c{}^a q_a X_4/(\rho h)\\
\rho h c_s^2 h^{ba}q_a X_b\\
(p/\rho)h^{ba} q_a X_b
\end{bmatrix}.
\label{eq:degen}
\end{align}
Three linearly independent solutions of Eq.\ \eqref{eq:degen}
are:
\begin{enumerate}
\item[(i)]
$X_b=0$, $X_4=0$, $X_5=1$.
\item[(ii)]
$X_4=X_5=0$, $X_b\sim \t_b$, where the ``tangential'' vector
$\t_b$ is orthogonal to $q^a$ and $u^a$.
(These vectors are tangential because $q^a=yu^a+\zeta^a$ implies
that they are spatial in the comoving frame and orthogonal to the normal
vector $\zeta^a$.)
One can find two such linearly independent tangential vectors
$\t_b^{(1)}$ and $\t_b^{(2)}$, and one can
choose them to be normalized and
orthogonal to each other.
\end{enumerate}
Since the normalization of the eigenvectors is arbitrary, we can take them
to be
\begin{equation}
X_1=\begin{bmatrix}
\t_b^{(1)}\\
0\\
0
\end{bmatrix},
\quad
X_2=\begin{bmatrix}
\t_b^{(2)}\\
0\\
0
\end{bmatrix},
\quad
X_3=\begin{bmatrix}
0\\
0\\
1
\end{bmatrix}.
\label{eq:eigendegen}
\end{equation}

\subsection{Right Eigenvectors for the Non-degenerate Eigenvalues}

In this case the eigenvectors satisfy
\begin{equation}
\begin{bmatrix}
u^a q_a h_c{}^b & h_c{}^a q_a/(\rho h) & 0\\[3pt]
\rho h c_s^2 h^{ba}q_a     & u^a q_a    & 0\\[3pt]
(p/\rho)h^{ba}q_a & 0    & u^a q_a
\end{bmatrix}
\begin{bmatrix}
X_b\\
X_4\\
X_5
\end{bmatrix}=0.
\label{eq:nondegen}
\end{equation}
Eliminating $h^{ba}q_a X_b$ from the second and third components of this
equation gives
\begin{equation}
X_5=\frac{p}{\rho^2 h c_s^2}X_4.
\end{equation}
The first component gives
\begin{equation}
u^a q_a h_c{}^b X_b+ h_c{}^a q_a/(\rho h)X_4=0.
\end{equation}
The 3 equations in
Eq.\ \eqref{eq:nondegen} are not linearly independent,
so we can regard $X_4$ as arbitrary.
 We get a simple form for
the eigenvectors
by taking $X_4=(\rho^2 h c_s^2/p) u^a q_a$, so that the eigenvectors are
\begin{equation}
X_\pm=
\begin{bmatrix}
-h_c{}^a q_a \rho c_s^2/p\\
(\rho^2 h c_s^2/p) u^a q_a\\
 u^a q_a
\end{bmatrix}.
\label{eq:eigennondegen}
\end{equation}
Here the two roots of the eigenvalue equation \eqref{eq:quadratic}
give two different values  $q^a_\pm$.

\subsection{Left Eigenvectors}

Left eigenvectors satisfy the equation
\begin{align}
\label{eq:lefteigen}
0&=L(A^a q_a)\\
&=\begin{bmatrix}
L^c & L_4 & L_5
\end{bmatrix}
\begin{bmatrix}
u^a q_a h_c{}^b & h_c{}^a q_a/(\rho h) & 0\\[3pt]
\rho h c_s^2 h^{ba}q_a     & u^a q_a    & 0\\[3pt]
(p/\rho)h^{ba}q_a & 0    & u^a q_a
\end{bmatrix}\notag\\
&=
\begin{bmatrix}
u^a q_a L^c h_c{}^b
+ h^{ba}q_a(\rho h c_s^2 L_4+ (p/\rho)L_5)\\[3pt] 
 h_c{}^a q_a L^c/(\rho h)
+ u^a q_a L_4\\[3pt]
u^a q_a L_5
\end{bmatrix}.
\label{eq:lefteigen2}
\end{align}

For the degenerate case ($u^a q_a=0$), this reduces to
\begin{equation}
0=
\begin{bmatrix}
\rho h c_s^2 h^{ba}q_a(\rho h c_s^2 L_4+ (p/\rho)L_5)\\[3pt]
 h_c{}^a q_a L^c/(\rho h)
\\[3pt]
0
\end{bmatrix}.
\end{equation}
Three linearly independent solutions are
\begin{equation}
\begin{split}
L_1 &= \begin{bmatrix}
\t_{(1)}^c & 0 & 0
\end{bmatrix},\\
L_2 &= \begin{bmatrix}
\t_{(2)}^c & 0 & 0
\end{bmatrix},\\
L_3 &= \begin{bmatrix}
0 & -p/(\rho^2 h c_s^2) & 1
\end{bmatrix}.
\end{split}
\label{eq:leftdegen}
\end{equation}

For the non-degenerate eigenvalues,
the first component of Eq.\ \eqref{eq:lefteigen2} shows that $L^c$ is
proportional to $h^c{}_bq^b$. This leaves three scalar equations that
are not linearly independent. Choosing $L_4$ to be
$q^a u_a$ gives
\begin{equation}
L_\pm = \begin{bmatrix}
-\rho h c_s^2 h^c{}_a q^a & q^a u_a & 0
\end{bmatrix}.
\label{eq:leftnondegen}
\end{equation}

\subsection{Orthogonality}
\label{I-sec:orthogonality}
Consider two different eigenvalues $y$ and $y'$, with $y\neq y'$. Then,
using Eq.\ \eqref{qalpha} in Eqs.\ \eqref{eq:eigenvalue}
and \eqref{eq:lefteigen} gives
\begin{align}
\label{eq:firstone}
A^a\zeta_a\, X &= -y A^a \xi_a\, X,\\
L'\, A^a\zeta_a &= -y' L'\, A^a \xi_a.
\label{eq:secondone}
\end{align}
Now make the usual
argument of multiplying Eq.\ \eqref{eq:firstone} on the left by $L'$
and Eq.\ \eqref{eq:secondone} on the right by $X$ and subtracting.
Thus deduce the orthogonality condition
\begin{equation}
L'\, A^a \xi_a\, X=0.
\end{equation}
Note that for eigenvectors corresponding to a degenerate eigenvalue
($y=y'$), one can show that a set
of left eigenvectors satisfying this mutual orthogonality condition
can always be chosen.

For the comoving eigenvectors we have considered so far,
the orthogonality condition holds with $\xi_a=u_a$:
\begin{equation}
L'\, A^a u_a\, X=0.
\label{eq:orthog}
\end{equation}
From Eq.\ \eqref{eq:amatrix}, we find
\begin{equation}
A^a u_a=
\begin{bmatrix}
-h_c{}^b & 0 & 0\\[3pt]
0   & -1    & 0\\[3pt]
0 & 0    & -1
\end{bmatrix},
\end{equation}
which is (minus) the identity matrix for our effectively
5-dimensional vectors and
so can be ignored in Eq.\ \eqref{eq:orthog}.

We can explicitly verify the orthogonality of the eigenvectors we have found.
Letting $\left[ L\right]$ denote the matrix of all 5 left row eigenvectors
and $\left[ X\right]$ the matrix of corresponding right column eigenvectors,
we have
\begin{widetext}
\begin{align}
\left[ L\right] \left[ X\right]&=
\begin{bmatrix}
\t^c_{(1,2)} & 0 & 0\\
0 & -p/(\rho^2 h c_s^2) & 1\\
-\rho h c_s^2 h^{ca} q^\pm_a & q^\pm_a u^a & 0
\end{bmatrix}
\begin{bmatrix}
\t_c^{(1,2)} & 0 & -h_c^a q^\pm_a (\rho c_s^2/p)\\
0 & 0 & (\rho^2 h c_s^2/p) q^\pm_a u^a\\
0 & 1 & q^\pm_a u^a
\end{bmatrix}\nonumber\\
&=
\begin{bmatrix}
1 & 0 & 0\\ 
0 & 1 & 0\\
0 & 0 & (\rho^2 h c_s^2/p)(c_s^2 h^{ab}q_a^\pm q_b^\pm+(u^a q_a^\pm)
(u^b q_b^\pm))
\end{bmatrix}.
\label{eq:orthog2}
\end{align}
\end{widetext}
%
Using Eqs.\ \eqref{eq:uq}, \eqref{eq:qqh} and
\eqref{eq:mucs}, we see that
the bottom right term evaluates to zero when the plus or minus signs
are opposite, showing orthogonality. (Note that Eq.\ \ref{eq:qqh}
holds for any pattern of plus and minus signs in $q_a$.)
When the signs are the same, the nonzero value can be used to renormalize the
left eigenvector so that the bottom right entry in Eq.\ \eqref{eq:orthog2}
becomes 1 (actually, a $2\times 2$ identity matrix).

\section{Changing Variables}
\label{sec:changing}

Having derived the characteristic decomposition with respect to some
set of variables,
we will find it useful to be able to change the variables and get the
decomposition in the new variables. For example,
in the previous sections we used the pair $(p,\epsilon)$ as the primitive
thermodynamic variables. We may wish to use a different pair, for example,
$(\rho,\epsilon)$, the most commonly used variables in numerical codes.
As another example,
we would like to use a set of conserved variables instead of
the set of primitive variables used above.

Let's consider the general problem of switching from a set of variables
$U_{\rm old}$ to a new set of variables $U$. We have
\begin{equation}
0=A^a_{\rm old}\nabla_a U_{\rm old},\quad
0=A^a\nabla_a U=A^a\frac{\partial U}{\partial U_{\rm old}}
  \nabla_a U_{\rm old}.
\end{equation}
Here $\partial U/\partial U_{\rm old}$ is the Jacobian matrix of the
transformation.
Thus we find
\begin{equation}
A^a_{\rm old}=A^a\frac{\partial U}{\partial U_{\rm old}}.
\label{eq:trans}
\end{equation}
We can therefore show that the transformation law for the
right eigenvectors is
\begin{equation}
X=\frac{\partial U}{\partial U_{\rm old}}X_{\rm old}
\label{eq:transeigen}
\end{equation}
and
verify the well-known result that the characteristic
eigenvalues are independent of the choice of variables.
For
\begin{multline}
0=(A^a_{\rm old}q_a)X_{\rm old}
= \left(A^a\frac{\partial U}{\partial U_{\rm old}}q_a\right)X_{\rm old}\\
= (A^a q_a)\frac{\partial U}{\partial U_{\rm old}}X_{\rm old}
=(A^a q_a)X,
\label{eq:trans_derivation}
\end{multline}
which shows that the transformation \eqref{eq:transeigen} provides the
eigenvectors of the transformed vector of matrices $A^a$
with the same quantities $q_a$, that is, the same eigenvalues $y$.
(Below we will discuss the case of changing $q^a$ as part of a transformation.)

An alternative method of changing variables is to go back to
the basic equations \eqref{eq:euler}, \eqref{eq:rhocons} and
\eqref{eq:epsilon} and formulate them in terms of the new variables.
Then one reads off the new $A^a$
and finds the eigenvectors. Generally, the transformation given in
Eq.\ \eqref{eq:transeigen} is much simpler and quicker.

\section{Primitive Variable Formulation Using the Eulerian Velocity}
\label{sec:eulerian}

Virtually all current hydrodynamics codes use a fluid 3-velocity instead of
the 4-velocity $u^a$ as a primitive variable.
We adopt the popular choice of the Eulerian velocity,
that is, the velocity measured by the observer whose 4-velocity
is $n^a$, the unit vector normal to the $t=\text{constant}$
hypersurface. This is the observer at rest in the 
$3+1$ coordinate system defined with respect to $n^a$.
In covariant form, this 3-velocity is defined as
\begin{align}
v^a&=\gamma^{ab}u_b/W\nonumber\\
&=u^a/W-n^a,
\label{eq:veuler}
\end{align}
where
\begin{align}
\label{eq:W}
W&\equiv -n_a u^a,\\
\gamma^{ab}&\equiv g^{ab}+n^a n^b.
\label{eq:project}
\end{align}
Here $\gamma^{ab}$ is the 3-metric introduced in
Eq.\ \eqref{eq:3+1}. Equivalently, it is the projection tensor
orthogonal to $n^a$ that produces purely spatial vectors.

Squaring Eq. \eqref{eq:veuler}, we see that
the quantity $W$ is a generalized gamma-factor:
\begin{equation}
W=\frac{1}{\sqrt{1-v^a v_a}}.
\label{eq:Wv}
\end{equation}
We can also write \eqref{eq:veuler} as
\begin{equation}
u^a=W(v^a+n^a),
\label{eq:uandv}
\end{equation}
which we recognize as the Lorentz transformation between
the time basis vectors $u^a$ and $n^a$ of the Lagrangian and Eulerian
frames.

Changing primitive variables from $u^a$ to $v^a$ is not simply a
transformation of variables. We are also implicitly changing the
timelike vector $\xi^a$ in Eq.\ \eqref{qalpha} defining the time
derivative operator from $u^a$ to $n^a$.
So the eigenvectors of the characteristic decomposition change
both because of the change of variables and also because of the
change from comoving to Eulerian time derivatives. We will need to deal with
both of these effects.

Now here comes a key point:
The change from comoving to Eulerian time derivatives can be incorporated
by changing $q^a$ from $y_u u^a+\zeta^a$ to $y n^a+s^a$, where $s^a$ is
a spatial unit normal vector in the Eulerian frame.
To implement this, we keep $X_{\rm old}$ as a function of $q^a$
and don't use the explicit representation of $q^a$ in terms of $u^a$.
Then \emph{after} the variable transformation
\eqref{eq:transeigen}
we can simply replace $q^a$ by its representation in terms of $n^a$.

\subsection{Transformation from $u^a$ to $v^a$}
As we will see, we will need both the Jacobian matrix for the
transformation from $u^a$ to $v^a$ and the matrix for
the inverse transformation from $v^a$ to $u^a$. We start with
the inverse transformation first, since it is slightly simpler.

To differentiate Eq.\ \eqref{eq:uandv} correctly,
write it as
\begin{equation}
u^a=W(\gamma^a{}_b v^b+n^a).
\end{equation}
Then
\begin{align}
\frac{\partial u^a}{\partial v^b}&=W \gamma^a{}_b+\frac{\partial W}{\partial
v^b}(v^a+n^a)\notag\\
&=W \gamma^a{}_b + W^2 u^a v_b,
\label{eq:dudv}
\end{align}
where we have differentiated Eq.\ \eqref{eq:Wv} to get the second line.

If the Jacobian matrix \eqref{eq:dudv} had full rank (i.e., rank 4), then
the matrix corresponding to its inverse transformation would simply be
the inverse matrix. However, because of the constraint $u^a u_a=-1$, the
rank is only 3, and we have to be more careful. 


\subsection{Transformation from $v^a$ to $u^a$}

To see how to proceed, work first in the local inertial frame of the Eulerian
observer, where we can use special relativity.
The fluid velocity has only 3 spatial components, $v^i$, and
$u^a=(W,Wv^i)$.
By explicit differentiation we get the $4\times 3$ matrix
\begin{equation}
\frac{\partial u^a}{\partial v^j}=
\begin{bmatrix}
W^3 v_j\\[2pt]
W \delta^k{}_j+W^3 v^k v_j
\end{bmatrix},
\label{eq:dudvflat}
\end{equation}
which is equivalent to Eq.\ \eqref{eq:dudv}.

For the inverse transformation, differentiate the expression
$v^i=u^i/u^0$ to get
the $3\times 4$ matrix
\begin{equation}
\frac{\partial v^i}{\partial u^a}=
\begin{bmatrix}
-\dfrac{v^i}{W} & \dfrac{1}{W}\delta^i{}_k
\end{bmatrix},
\label{eq:dvduSR}
\end{equation}
Even though these matrices are not inverses, it is still true that
\begin{equation}
\frac{\partial v^i}{\partial u^a}
\frac{\partial u^a}{\partial v^j}=\delta^i{}_j.
\label{eq:inverse}
\end{equation}
This is a special case of a general
theorem\cite{strang2006}: If a matrix $P$ is $m\times n$
with $m < n$, and $P$ has full rank, i.e., $\rank(P)=m$, then
$P$ has a right inverse.
This is an $n\times m$ matrix $Q$ such that $P\cdot Q=1$.
Note that the inverse is not unique.
One can also state the theorem for the case $m>n$ with the rank of the matrix
equal to $n$, in which case the inverse is a left inverse: $Q\cdot P=1$.
This
is the case we are dealing with, with
$Q=\partial v^i/\partial u^a$ and $P=\partial u^a/\partial v^j$.

We now want to generalize this procedure to obtain the covariant
expression  $\partial v^a/\partial u^b$ valid in any coordinate system.
The most general expression for this quantity can be written in terms of
$n^a$, $u^a$, and the metric:
\begin{equation}
\frac{\partial v^a}{\partial u^b}=
A\delta^a{}_b+Bu^a u_b + C u^a n_b + D n^a u_b + En^a n_b.
\label{eq:utrans1}
\end{equation}
There is no need to allow for terms involving
$v^a$ because of Eq.\ \eqref{eq:uandv}.
In a general coordinate system, the 3-dimensional inverse condition
\eqref{eq:inverse} is expressed using the 3-dimensional spatial metric:
\begin{equation}
\frac{\partial v^a}{\partial u^c}
\frac{\partial u^c}{\partial v^b}=\gamma^a{}_b.
\label{eq:inverse2}
\end{equation}
Substituting Eqs.\ \eqref{eq:dudv} and \eqref{eq:utrans1} in
Eq.\ \eqref{eq:inverse2} gives
\begin{equation}
A=1/W,\quad C=1/W^2,\quad E=0.
\end{equation}

The coefficients $B$ and $D$ remain arbitrary, reflecting the non-uniqueness
of the inverse. 
We will fix  this non-uniqueness by imposing an additional crucial requirement.
(We will see below why this requirement is important.)
Even though no theorem guarantees that we can do so, we are
able to impose the matrix ``inverse'' requirement
\begin{equation}
\frac{\partial u^a}{\partial v^c}
\frac{\partial v^c}{\partial u^b}=h^a{}_b.
\label{eq:inverse3}
\end{equation}
We find that this condition is satisfied provided we take $B=0$.
The coefficient $D$ remains arbitrary. In our applications,
$\partial v^a/\partial u^b$ always acts on quantities that are spatial
with respect to $n^a$ or $u^b$, so the $D$ term never contributes.
Accordingly, we can simply set $D=0$ and have
\begin{equation}
\frac{\partial v^a}{\partial u^b}
=\frac{1}{W}\delta^a{}_b+\frac{1}{W^2}u^a n_b.
\label{eq:vtrans}
\end{equation}
It is easy to check that this reduces to Eq.\ \eqref{eq:dvduSR}
in special relativity.

You can see the non-uniqueness of the inverse in the special relativistic
case above by, for example, differentiating the expression
$u^i/(1+u^j u_j)^{1/2}$, which is equivalent to $v^i=u^i/u^0$, the
expression we previously used.
You get a completely different expression for
Eq.\ \eqref{eq:dvduSR}, but it still satisfies Eq. \eqref{eq:inverse}.
However, it does not satisfy Eq.\ \eqref{eq:inverse3}.

\section{Transformation of Left Eigenvectors}
\label{sec:lefttrans}

For left eigenvectors, the transformation properties can be
somewhat more complicated than for right eigenvectors.
We need to consider separately
the cases where the transformation is invertible and where it is not.

\subsection{Invertible Transformation}

This case is straightforward. If we
write the equation $A^a \nabla_a U=0$ for the ``old'' variables
in some $3+1$ decomposition, we get
\begin{equation}
A^0_{\rm old}\frac{\partial U_{\rm old}}{\partial t}+
 A^i_{\rm old}\frac{\partial U_{\rm old}}{\partial x^i}=0.
\label{eq:standard}
\end{equation}
As usual, since we deal only with the principal part of the equations
in this paper, we can ignore extra terms from, for example, connection
coefficients, since they do not contain derivatives of the dependent variables.
Multiply Eq.\ \eqref{eq:standard} by $(A^0_{\rm old})^{-1}$ to get
\begin{equation}
\frac{\partial U_{\rm old}}{\partial t}+
\left(A^0_{\rm old}\right)^{-1}
 A^i_{\rm old}\frac{\partial U_{\rm old}}{\partial x^i}=0.
\label{eq:leftold}
\end{equation}
If we take the vector $q_a$ in Eq.\ \eqref{qalpha} to be
$q_a=y n_a+s_a$, then the left eigenvector problem for Eq.\ \eqref{eq:leftold}
is
\begin{equation}
-y L_{\rm old}+L_{\rm old}\left(A^0_{\rm old}\right)^{-1}A^i_{\rm old}s_i=0.
\label{eq:lold1}
\end{equation}
Now since the transformation is invertible, Eq.\ \eqref{eq:trans} gives
\begin{equation}
\left(A^a_{\rm old}\right)^{-1}=\frac{\partial U_{\rm old}}{\partial U}
\left(A^a\right)^{-1}.
\label{eq:inversetrans}
\end{equation}
Using this equation for $\left(A^0_{\rm old}\right)^{-1}$ in
Eq.\ \eqref{eq:lold1} gives
\begin{equation}
-y L_{\rm old}\frac{\partial U_{\rm old}}{\partial U}
+L_{\rm old}\frac{\partial U_{\rm old}}{\partial U}
\left(A^0\right)^{-1}A^i s_i=0.
\end{equation}
Thus
\begin{equation}
L=L_{\rm old}\frac{\partial U_{\rm old}}{\partial U}=
L_{\rm old}\left(\frac{\partial U}{\partial U_{\rm old}}\right)^{-1},
\label{eq:invertible}
\end{equation}
as expected. Since the coefficient of $\partial/\partial t$ in
Eq.\ \eqref{eq:leftold} is unity, we have the simple orthogonality
conditions
\begin{equation}
\left[ L_{\rm old}\right] \left[ X_{\rm old}\right]=
\left[ L\right] \left[ X\right]=\one,
\end{equation}
where $\one$ denotes the identity matrix.

\subsection{Quasi-invertible Transformation}
\label{sec:quasi}
We will call a transformation \emph{quasi-invertible} if
\begin{equation}
A^a_{\rm old}=A^a\frac{\partial U}{\partial U_{\rm old}}
\quad \Leftrightarrow\quad
A^a=A^a_{\rm old}\frac{\partial U_{\rm old}}{\partial U}.
\label{eq:quasiinvertible}
\end{equation}
The simplest way to satisfy the quasi-invertibility condition
is if
\begin{equation}
\frac{\partial U_{\rm old}}{\partial U}
\frac{\partial U}{\partial U_{\rm old}}=\one,
\quad
\frac{\partial U}{\partial U_{\rm old}}
\frac{\partial U_{\rm old}}{\partial U}=\one,
\label{eq:simpleqi}
\end{equation}
but we will see in Paper II that a more general relation is needed for GRMHD.
To keep the current paper self-contained, we give a preview of this
generalization in Appendix \ref{app:quasiinv}.

The transformation from $u^a$ to $v^a$ is an example of a quasi-invertible
transformation, as shown by the relations
\eqref{eq:inverse2} and \eqref{eq:inverse3}.
(The full transformation matrices are the velocity parts \eqref{eq:dudv}
and \eqref{eq:vtrans} extended with $2\times 2$ identity matrices for
the thermodynamic variables.)

We now determine the transformation
of left eigenvectors under this transformation. Let $\widetilde L$ denote
a tentative new eigenvector satisfying
\begin{align}
0&=\widetilde L(A^a q_a)\notag\\
&=\widetilde L \left(A^a_{\rm old}
  \frac{\partial U_{\rm old}}{\partial U}\right)q_a
  \quad\text{(by Eq.\ \ref{eq:quasiinvertible})}\notag\\
&=
\widetilde L\left(A^a_{\rm old}q_a\right)
  \frac{\partial U_{\rm old}}{\partial U}.
\end{align}
Comparing with $L_{\rm old}(A^a_{\rm old}q_a)=0$, we see that we
can take $\widetilde L = L_{\rm old}$. But when we carry out the
transformation from $u^a$ to $v^a$, we also want to change
from a $3+1$ decomposition with respect to $u^a$ to one
with respect to $n^a$.
So when we set $\widetilde L = L_{\rm old}$, we keep the functional
dependence on $q^a$ while replacing $q^a=y_u u^a+
\zeta^a$ by $q^a = y n^a+s^a$. This is the same procedure described
above for
transforming the right eigenvectors. Moreover,
the orthogonality condition changes
from
\begin{equation}
[L_{\rm old}]\left(A^a_{\rm old}u_a\right)[X_{\rm old}]=\one
\end{equation}
(cf.\ Eq.\ \ref{eq:orthog}) to
\begin{equation}
[\widetilde L]\left(A^a_{\rm old}n_a\right)\frac{\partial U_{\rm old}}{\partial U}
[X]=\one.
\end{equation}
Since $\widetilde L=L_{\rm old}$, this gives
\begin{equation}
L=L_{\rm old}\left(A^a_{\rm old}n_a\right)
  \frac{\partial U_{\rm old}}{\partial U},
\label{eq:ltrans}
\end{equation}
with the understanding about updating $q^a$ in $L_{\rm old}$.
The matrix of left eigenvectors $[L]$ will then be the matrix inverse
of $[X]$, as desired in computational work.

\section{Eigenvectors in the Eulerian frame}
\label{sec:eulerianframe}

\subsection{Eigenvalues for the Eulerian frame}

The triply degenerate eigenvalue is given by
\begin{equation}
0= u^a q_a=W(v^a+n^a)(y n_a+s_a)=W(v_n-y),
\label{eq:qu}
\end{equation}
or $y=v_n$.
Here $v_n$ denotes the normal component of the velocity, $v^a s_a$.

The non-degenerate eigenvalues are given by Eq.\ \eqref{eq:quadratic}.
Substituting $h^{ab}=g^{ab}+u^a u^b$, this equation
becomes
\begin{equation}
(u^a q_a)^2-c_s^2[q_a q^a+(u^a q_a)^2]=0.
\label{eq:quadratic2}
\end{equation}
Using
\begin{equation}
q_a q^a=1-y^2
\label{eq:qq}
\end{equation}
and $u^a q_a=W(v_n-y)$ from Eq.\ \eqref{eq:qu}
in Eq.\ \eqref{eq:quadratic2}, we get a quadratic equation in $y$ with roots
\begin{align}
\label{eq:roots1}
y_\pm&=\frac{(1-c_s^2)W^2v_n\pm c_s d}{W^2+c_s^2(1-W^2)}\\
&=\frac{(1-c_s^2)v_n\pm c_s d/W^2}
  {1-v^2c_s^2},
\label{eq:roots}
\end{align}
where
\begin{align}
d&\equiv \sqrt{c_s^2+(1-c_s^2)(1-v_n^2)W^2}\\
&=W\sqrt{1-v^2c_s^2-v_n^2(1-c_s^2)}.
\end{align}
In these expressions, the second form is the preferred computational
form.
Equation \eqref{eq:roots}
is also given in Ref.\cite{schoepe2018}.
Equation \eqref{eq:roots}
is the generalization to an arbitrary unit
normal of the well-known
formula~\cite{banyuls1997,ibanez2001,font2008}
in which the spatial normal vector is unnormalized with components
$(1,0,0)$.

The eigenvalues satisfy the following relations:
\begin{equation}
\begin{split}
\frac{1}{y_{\pm}-v_n} &= \frac{\pm d+c_s v_n}{c_s(1-v_n^2)},\\
\frac{y_{\pm}}{y_{\pm}-v_n} &= \frac{c_s \pm  v_n d}{c_s(1-v_n^2)},\\
\frac{1-y_\pm v_n}{y_\pm-v_n}&= \pm\frac{d}{c_s}.
\end{split}
\label{eq:eig_idents}
\end{equation}
These relations are useful for expressing eigenvectors in alternative
forms and comparing results with other authors.

\subsection{Right Eigenvectors in the Eulerian frame}

Right eigenvectors transform according to \eqref{eq:transeigen}.
We have to be careful dealing with the
tangential vectors $\t^a_{(1)}$ and $\t^a_{(2)}$ in the eigenvectors
$X_1$ and $X_2$ of Eq.\ \eqref{eq:eigendegen}.
When transforming eigenvectors, we have to keep the same functional
dependence on $q^a$. But
recall from the discussion below Eq.\ \eqref{eq:degen}
that $\t^a_{(1)}$ and $\t^a_{(2)}$ are required to be orthogonal to
$q^a$ and $u^a$, that is, to $u^a$ and the spatial normal $\zeta^a$.
However, when we update $q^a$ to be $y n^a + s^a$, we now need tangential
vectors orthogonal to $n^a$ and $s^a$. Denote
the new tangent vectors by
the symbols $t^a_{(1,2)}$ in ordinary font.
If we apply the transformation \eqref{eq:vtrans} to these vectors,
they remain unchanged up to a normalization factor, which we can ignore.
The remaining degenerate eigenvector $X_3$
is unchanged because it has a zero vector in its first entry.
Using $\xbf$ to denote eigenvectors in the Eulerian frame, we have
\begin{equation}
\xbf_1=\begin{bmatrix}
t^a_{(1)}\\
0\\
0
\end{bmatrix},
\quad
\xbf_2=\begin{bmatrix}
t^a_{(2)}\\
0\\
0
\end{bmatrix},
\quad
\xbf_3=\begin{bmatrix}
0\\
0\\
1
\end{bmatrix}.
\end{equation}

For the eigenvectors $X_\pm$ of Eq.\ \eqref{eq:eigennondegen},
the first element
transforms to
\begin{align}
&-\left(\frac{1}{W}\delta^b{}_c+\frac{1}{W^2}u^b n_c\right)
q_a h^{ac}\rho c_s^2/p\nonumber\\
=&-\left(\frac{1}{W}q^b+\frac{1}{W^2}u^b n^c q_c\right)
 \rho c_s^2/p\nonumber\\
=&-\frac{1}{W}\left(\gamma^{b c}q_c+v^b n^c q_c\right)
 \rho c_s^2/p.
\end{align}
Renormalizing by the factor $-\rho c_s^2/(W p)$, we get
\begin{equation}
\xbf_\pm=
\begin{bmatrix}
\gamma^{b c}q_c+v^b n^c q_c\\[3pt]
-\rho h  W u^a q_a\\[3pt]
-(p/\rho c_s^2)W u^a q_a
\end{bmatrix}.
\label{eq:xpmv}
\end{equation}
As expected, the first entry in the eigenvector is completely spatial.

To complete the transformation to the Eulerian frame, we must evaluate
Eq.\ \eqref{eq:xpmv} with $q^a$ defined in that frame: $q^a=y n^a+s^a$.
We get
\begin{equation}
\xbf_\pm=
\begin{bmatrix}
s^i-y_\pm v^i\\
-\rho h W^2 (v_n-y_\pm)\\
-p W^2 (v_n-y_\pm)/(\rho c_s^2)
\end{bmatrix}.
\end{equation}

\subsection{Left Eigenvectors in the Eulerian frame}

For the left eigenvectors, we use Eq.\ \eqref{eq:ltrans}. Using
Eqs.\ \eqref{eq:amatrix} and \eqref{eq:dudv}, we get
\begin{widetext}
\begin{align}
\left(A^a_{\rm old}n_a\right)\frac{\partial U_{\rm old}}{\partial U}
&=
\begin{bmatrix}
-W h_c{}^d & n_a h_c{}^a/(\rho h) & 0\\[3pt]
\rho h c_s^2 n_a h^{da}     &  -W   & 0\\[3pt]
(p/\rho)n_a h^{da} & 0    & -W
\end{bmatrix}
\begin{bmatrix}
W \gamma_d{}^b + W^2 u_d v^b & 0 & 0\\[3pt]
0 & 1 & 0\\[3pt]
0 & 0 & 1
\end{bmatrix}
\notag\\
&= \begin{bmatrix}
-W^2 h_c{}^d \gamma_d{}^b & n_a h_c{}^a/(\rho h) & 0\\[3pt]
-\rho h c_s^2 W^3 v^b & -W & 0\\[3pt]
-(p/\rho)W^3 v^b & 0 & -W
\end{bmatrix}.
\end{align}
Now multiply on the left by the matrix $[L_{\rm old}]$, which appears
in Eq.\ \eqref{eq:orthog2}.
Just as in the case of the right eigenvectors discussed above,
the tangential vectors $\t^c_{(1,2)}$ appropriate for the
comoving case must be replaced by
the tangential vectors $t^c_{(1,2)}$ appropriate for the Eulerian case.
So we get
\begin{align}
[\lbf]&=\begin{bmatrix}
t^c_{(1,2)} & 0 & 0\\[3pt]
0 & -p/(\rho^2 h c_s^2) & 1\\[3pt]
-\rho h c_s^2 h^{ca} q^\pm_a & q^\pm_a u^a & 0
\end{bmatrix}
\begin{bmatrix}
-W^2 h_c{}^d \gamma_d{}^b & n_a h_c{}^a/(\rho h) & 0\\[3pt]
-\rho h c_s^2 W^3 v^b & -W & 0\\[3pt]
-(p/\rho)W^3 v^b & 0 & -W
\end{bmatrix}
\notag\\
&=
\begin{bmatrix}
-W^2 t^b_{(1,2)} -W^4 v_{1,2}v^b & -v_{1,2}W^2/(\rho h) & 0\\[3pt]
0 & pW/(\rho^2 h c_s^2) & -W\\[3pt]
\rho h c_s^2 W^2 \gamma^b{}_a q^a & -c_s^2 n_a q^a-(1-c_s^2)W q_a u^a & 0
\end{bmatrix},
\label{eq:leuler}
\end{align}
\end{widetext}
where we have carried out some elementary simplifications in the second
equality.
Here $v_1=v_a t^a_{(1)}$ and similarly for $v_2$.

If we rescale the
degenerate eigenvectors in the
first row of Eq.\ \eqref{eq:leuler} by $-W^2$, we get
\begin{align}
\lbf_1&=
\begin{bmatrix}
t^b_{(1)} +W^2 v_1v^b & v_1/(\rho h) & 0
\end{bmatrix},\\
\lbf_2&=
\begin{bmatrix}
t^b_{(2)} +W^2 v_2v^b & v_2/(\rho h) & 0
\end{bmatrix}.
\end{align}
We can get alternative expressions for these eigenvectors by forming
linear combinations, which is valid since the eigenvectors span a
degenerate subspace.
Write
\begin{equation}
v^b=v_1 t^b_{(1)}+v_2 t^b_{(2)}+v_n s^b
\end{equation}
in each of these expressions.
Next, eliminate $t^b_{(2)}$ from the expressions. This gives
\begin{equation}
\begin{bmatrix}
v_1 v_n s^b+(v_1^2+v_2^2+1/W^2)t^b_{(1)} &  v_1/(\rho h W^2) & 0
\end{bmatrix}.
\end{equation}
Finally, replace $v_1^2+v_2^2$ by $v^2-v_n^2=1-1/W^2-v_n^2$. Interchanging
$1$ and $2$ gives the other eigenvector. We find
\begin{equation}
\lbf_{1,2}=
\begin{bmatrix}
v_{1,2} v_n s^b+(1-v_n^2)t^b_{(1,2)} & v_{1,2}/(\rho h W^2) & 0
\end{bmatrix}.
\end{equation}

For the non-degenerate eigenvectors (last row in Eq. \ref{eq:leuler}),
if we substitute for $q^a$ we get
\begin{align}
&\lbf_\pm=\notag\\
&\begin{bmatrix}
\rho h c_s^2 W^2 s^i &
y_\pm [W^2+c_s^2(1-W^2)]-(1-c_s^2)W^2v_n & 0
\end{bmatrix}\notag\\
&=
\begin{bmatrix}
\rho h c_s^2 W^2 s^i &
\pm c_s d & 0
\end{bmatrix},
\end{align}
where we have used Eq.\ \eqref{eq:roots1}.
To make contact with previous literature, note that using
Eq.\ \eqref{eq:eig_idents}, this expression can
also be written as
\begin{equation}
\lbf_\pm=
\begin{bmatrix}
\rho h c_s^2 W^2 s^i &
\dfrac{c_s^2(1-v_ny_\pm)}{y_\pm-v_n} & 0
\end{bmatrix}.
\end{equation}

\subsection{Transformation to $(\rho,\epsilon)$}

As mentioned previously, most numerical codes use the pair
$(\rho,\epsilon)$ as primitive variables. Transforming from
$(p,\epsilon)$ to $(\rho,\epsilon)$ is a simple example of
an invertible transformation:
\begin{align}
\label{eq:kappa}
\frac{dU_{\rm old}}{dU}&=
\frac{\partial(p,\epsilon)}{\partial(\rho,\epsilon)}
=\begin{bmatrix}
\chi & \kappa\\
0 & 1
\end{bmatrix},\\
\frac{dU}{dU_{\rm old}}&=
\frac{\partial(\rho,\epsilon)}{\partial(p,\epsilon)}
=\begin{bmatrix}
1/\chi & -\kappa/\chi\\
0 & 1
\end{bmatrix}.
\label{eq:inversekappa}
\end{align}
The part of the transformation acting on the vector part of the eigenvectors
is the identity matrix, so we have not bothered to write it explicitly.
Applying Eq.\ \eqref{eq:inversekappa} to the Eulerian right eigenvectors above,
we get
\begin{widetext}
\begin{equation}
\xbf_{1,2}=\begin{bmatrix}
t^i_{(1,2)}\\
0\\
0
\end{bmatrix},
\quad
\xbf_3=\begin{bmatrix}
0\\
-\kappa\\
\chi
\end{bmatrix},
\quad
\xbf_\pm=
\begin{bmatrix}
s^i-y_\pm v^i\\
-\rho W^2 (v_n-y_\pm)/c_s^2\\
-p W^2 (v_n-y_\pm)/(\rho c_s^2)
\end{bmatrix}.
\label{eq:fonteigen}
\end{equation}
Similarly, applying Eq.\ \eqref{eq:kappa} to the left eigenvectors gives
\begin{equation}
\begin{split}
\lbf_{1,2}&=\frac{1}{1-v_n^2}\begin{bmatrix}
(1-v_n^2)t_i^{(1,2)}+v_n v_{1,2} s_i
&
\dfrac{\chi v_{1,2}}{\rho h W^2}
&
\dfrac{\kappa v_{1,2}}{\rho h W^2}
\end{bmatrix},\\
\lbf_3&=\frac{1}{h c_s^2}\begin{bmatrix}
0 &
-\dfrac{p}{\rho^2} & 1\end{bmatrix},\\
\lbf_\pm&=\pm\frac{1}{(1-v_n^2)(y_- - y_+)}
\begin{bmatrix}
(y_\mp-v_n)s_i &
-\dfrac{(1-y_\mp v_n)\chi}{\rho h W^2} &
-\dfrac{(1-y_\mp v_n)\kappa}{\rho h W^2}
\end{bmatrix}.
\end{split}
\label{eq:leftfontorig}
\end{equation}
\end{widetext}
Here we have normalized the eigenvectors so that $[\lbf][\xbf]=\one$.

Using the relations \eqref{eq:eig_idents},
we can rewrite the non-degenerate eigenvectors in the alternative form
\begin{align}
\label{eq:altright}
&\xbf_{(\pm)}=\notag\\
&\begin{bmatrix}
\dfrac{1}{1-v_n^2}[(c_sv_n\pm d)s^i-(c_s\pm v_n d)v^i]
&
\dfrac{\rho W^2}{c_s}
&
\dfrac{p  W^2}{\rho c_s}
\end{bmatrix}^T,\\
&\lbf_\pm=\frac{1}{2}
\begin{bmatrix}
\pm\dfrac{1}{d}s_i &
\dfrac{\chi}{\rho h W^2 c_s} &
\dfrac{\kappa}{\rho h W^2 c_s}
\end{bmatrix}.
\label{eq:altleft}
\end{align}
Here we have renormalized the eigenvectors by dividing $\xbf_\pm$ by
$(1-v_n^2)/(c_s v_n\pm d)$ and multiplying $\lbf_\pm$ by this same
factor.
These expressions agree with those in \cite{schoepe2018} if we transform
our expressions back to $(p,\epsilon)$.

Note that we have not bothered to introduce a separate
notation to distinguish $(p,\epsilon)$ eigenvectors from $(\rho,\epsilon)$
eigenvectors since these are only being used as intermediate
quantities on the way to using conserved variables.

\section{Conservative Formulation}
\label{sec:conservative}

The usual set of conserved variables for relativistic hydrodynamics is
\begin{equation}
U=(S^a,D,\tau),
\end{equation}
\vskip -\belowdisplayskip
\vskip \belowdisplayshortskip 
\noindent where
\vspace{-\abovedisplayskip}
\vspace{\abovedisplayshortskip} 
\begin{align}
S^a &= \rho h W^2 v^a,\\
D &= \rho W,\\
\tau &= \rho h W^2 - p -\rho W.
\end{align}
Note that there are only actually 5 independent variables
in the Eulerian frame since $S^0=0$.

\subsection{Transformation Matrix for Right Eigenvectors}
We get the eigenvectors for the conservative formulation by transforming
from a primitive variable formulation rather than deriving them
directly from the equations of motion in conservation form.
A convenient choice
is the variables we used in the previous section,
$U_{\rm old}=(v^b,\rho,\epsilon)$.
The transformation matrix
$\partial U/\partial U_{\rm old}$
consists of the following derivatives:
\begin{widetext}
\begin{equation}
\frac{\partial(S^a,D,\tau)}{\partial(v^b,\rho,\epsilon)}=
\begin{bmatrix}
\rho h W^2(\delta^a{}_b+2W^2v^a v_b) &
  (1+\epsilon+\chi)W^2 v^a &
  (\rho+\kappa)W^2 v^a\\[2pt]
\rho W^3 v_b &
W &
0\\[2pt]
\rho (2h W-1) W^3 v_b &
(1+\epsilon+\chi)W^2-\chi-W &
(\rho+\kappa)W^2-\kappa
\end{bmatrix}.
\label{eq:transtocons}
\end{equation}
\end{widetext}
Here the derivatives of $\rho h$ have been
computed using $\rho h = \rho+\rho\epsilon+p$.
It is convenient when using the matrix \eqref{eq:transtocons}
to replace $1+\epsilon$ by $h-p/\rho$ since this makes it easier to
introduce the sound speed \eqref{eq:cs}.

\subsection{Transformation Matrix for Left Eigenvectors}
The transformation \eqref{eq:transtocons}
is invertible. We need the inverse to compute
the left eigenvectors according to Eq.\ \eqref{eq:invertible}. There
are several ways to compute the inverse. One method is to use the
formulas for the inverse of a partitioned matrix.
However, this method is extremely complicated for the GRMHD case.
So instead we will use a method that is also tractable for GRMHD.
The method is based on an idea in
Refs.\ \cite{antonthesis,anton2010},
namely to compute the inverse transformation to \eqref{eq:transtocons}
by computing the necessary partial derivatives directly, using
an intermediate variable
\begin{equation}
Z=\rho h W^2=\tau + D + p.
\label{eq:zdef}
\end{equation}

We first compute the derivatives of the primitive variables
$(v^a,\rho,\epsilon)$ with respect to $\tau$, keeping $S^a$ and $D$ constant.
We have
\begin{equation}
\frac{\partial v^a}{\partial \tau}=\frac{\partial}{\partial \tau}
 \left(\frac{S^a}{Z}\right)=-\frac{S^a}{Z^2}\frac{\partial Z}{\partial \tau}
 =-\frac{v^a}{Z}\frac{\partial Z}{\partial \tau}.
\end{equation}
Thus Eq.\ \eqref{eq:Wv} gives
\begin{equation}
\frac{\partial W}{\partial \tau}= W^3 v_a \frac{\partial v^a}{\partial \tau}
=-\frac{W(W^2-1)}{Z}\frac{\partial Z}{\partial \tau}.
\end{equation}
Next,
\begin{equation}
\frac{\partial \rho}{\partial \tau}=\frac{\partial}{\partial \tau}
 \left(\frac{D}{W}\right)=(W^2-1)\frac{\rho}{Z}\frac{\partial Z}{\partial \tau}.
\label{eq:drhodtau}
\end{equation}
Finally, since
\begin{equation}
\epsilon=h-1-\frac{p}{\rho}=\frac{Z}{DW}-1-\frac{p}{\rho},
\end{equation}
we get after some simplifications
\begin{equation}
\frac{\partial \epsilon}{\partial \tau}=
\frac{1}{\rho}+\frac{p}{\rho}\frac{W^2-1}{Z}\frac{\partial Z}{\partial \tau}.
\label{eq:depsdtau}
\end{equation}
Here we computed the term $\partial p/\partial \tau$ from Eq.\ \eqref{eq:zdef}:
\begin{equation}
\frac{\partial Z}{\partial \tau}=1+\frac{\partial p}{\partial \tau}.
\label{eq:dpdtau}
\end{equation}

We now go beyond the method in Refs.\ \cite{antonthesis,anton2010}
by expressing the dependence on the equation of state $p=p(\rho,\epsilon)$
explicitly to get a final expression for
$\partial Z/\partial \tau$. We get this from Eq.\ \eqref{eq:dpdtau} using
Eq.\ \eqref{eq:chi}:
\begin{equation}
\frac{\partial Z}{\partial \tau}=1+\chi\frac{\partial \rho}{\partial \tau}
+\kappa\frac{\partial \epsilon}{\partial \tau}.
\end{equation}
We substitute from Eqs.\ \eqref{eq:drhodtau} and \eqref{eq:depsdtau} for
$\partial \rho/\partial \tau$ and $\partial \epsilon/\partial \tau$ and
solve the resulting linear equation for $\partial Z/\partial \tau$.
This gives
\begin{equation}
\frac{\partial Z}{\partial \tau}=
\frac{(\rho+\kappa)W^2}{\rho[W^2-c_s^2(W^2-1)]},
\end{equation}
where we have simplified the denominator using Eq.\ \eqref{eq:cs}.

The derivatives of the primitive variables with respect to $D$ and $S^a$
are computed in the same way. We find
\begin{widetext}
\begin{equation}
\frac{\partial(v^b,\rho,\epsilon)}{\partial(S^a,D,\tau)}=
\begin{bmatrix}
\dfrac{1}{Z}\delta^b{}_a-\dfrac{v^b}{Z}\pd{Z}{S^a} & -\dfrac{v^b}{Z}\pd{Z}{D}
  & -\dfrac{v^b}{Z}\pd{Z}{\tau}\\[7pt]
-\dfrac{\rho W^2}{Z}v_a+\dfrac{\rho(W^2-1)}{Z}\pd{Z}{S^a} &
 \dfrac{1}{W}+\dfrac{\rho(W^2-1)}{Z}\pd{Z}{D} &
\dfrac{\rho(W^2-1)}{Z}\pd{Z}{\tau}\\[7pt]
-\left(\dfrac{1}{\rho}+\dfrac{pW^2}{\rho Z}\right)v_a
  +\dfrac{p(W^2-1)}{\rho Z}\pd{Z}{S^a} &
\dfrac{p}{\rho^2 W}+\dfrac{1}{\rho}-\dfrac{h}{\rho W}
   +\dfrac{p(W^2-1)}{\rho Z}\pd{Z}{D} &
\dfrac{1}{\rho}+\dfrac{p(W^2-1)}{\rho Z}\pd{Z}{\tau}
\end{bmatrix},
\label{eq:inversecons}
\end{equation}
\end{widetext}
where
\begin{align}
\label{eq:dzds}
\pd{Z}{S^a}&=-\frac{(\kappa+\rho c_s^2)W^2}{\rho [W^2-c_s^2(W^2-1)]}v_a,\\
\label{eq:dzdD}
\pd{Z}{D}&=\frac{(\rho+\kappa)W^2+h W(\rho c_s^2-\kappa)}
  {\rho [W^2-c_s^2(W^2-1)]},\\
\pd{Z}{\tau}&=\frac{(\rho+\kappa)W^2}{\rho [W^2-c_s^2(W^2-1)]}.
\label{eq:dzdtau}
\end{align}
Although it's not needed in this paper, we give the explicit expression for
the inverse matrix \eqref{eq:inversecons} with the substitutions
\eqref{eq:dzds} -- \eqref{eq:dzdtau} in Appendix \ref{app:inverse}.

\subsection{Conservative Right Eigenvectors}

We use Eq.\ \eqref{eq:transeigen} with the matrix \eqref{eq:transtocons}
and the eigenvectors \eqref{eq:fonteigen} to get the right eigenvectors for
the conserved variables in the Eulerian frame.
Using $\Rbf$ to denote conservative right
eigenvectors, we find
\begin{widetext}
\begin{equation}
\Rbf_{1,2}=
\begin{bmatrix}
h(t^{(1,2)}_i+2W^2v_{1,2}v_i)\\[5pt]
W v_{1,2}\\[5pt]
W(2hW-1)v_{1,2}
\end{bmatrix},
\quad
\Rbf_{3}=
\begin{bmatrix}
hW(\kappa-\rho c_s^2)v_i\\[5pt]
\kappa\\[5pt]
hW(\kappa-\rho c_s^2)-\kappa
\end{bmatrix},
\quad
\Rbf_\pm=
\begin{bmatrix}
hW\left(v_i \pm\dfrac{c_s}{d}s_i\right)\\[5pt]
1\\[3pt]
h W\left(1\pm\dfrac{c_sv_n}{d}\right) -1
\end{bmatrix}.
\label{eq:consright}
\end{equation}
\end{widetext}

We have renormalized the eigenvectors slightly by some common factors, which
we will undo in the corresponding left eigenvectors to maintain the
overall normalization to be unity. These eigenvectors agree with those
in \cite{font2008}, but are somewhat simpler and also valid for any direction
of the normal vector $s^i$. Note that there does not seem to be any
advantage in transforming Eq.\ \eqref{eq:altright} to try to get a simpler
expression than in Eq.\ \eqref{eq:consright}.

Note that the expressions given in Eq.\ \eqref{eq:consright} remain
valid for barotropic equations of state, where $\kappa-\rho c_s^2\to 0$
(see Ref.~\cite{ibanez2012} and Appendix~\ref{app:c}).

\subsection{Conservative Left Eigenvectors}

Using Eqs.\ \eqref{eq:leftfontorig} and \eqref{eq:inversecons}
in the transformation equation \eqref{eq:invertible}, and restoring
the scale factors removed from the right eigenvectors, we get
\begin{equation}
\begin{split}
\Lbf_{1,2}&=\frac{1}{h (1-v_n^2)}
\left[
\begin{matrix}
v_{1,2}\,v_n s_i+(1-v_n^2) t^{(1,2)}_i\\
-v_{1,2}\\
-v_{1,2}
\end{matrix}
\right]^T,\\
\Lbf_{3}&=\frac{1}{\rho h c_s^2}
\left[
\begin{matrix}
W v_i\\[1pt]
h-W\\
-W
\end{matrix}
\right]^T,\\
\Lbf_{\pm}&=\frac{1}{2\rho h W c_s^2 (1-v_n^2)}
\left[
\begin{matrix}
-a v_i+ \rho c_s(c_s v_n\pm d) s_i\\[2pt]
b_\pm-h W(\kappa-\rho c_s^2)(1-v_n^2)\\
b_\pm
\end{matrix}
\right]^T,
\end{split}
\label{eq:left_cons}
\end{equation}
where
\begin{align}
a&=W^2(1-v_n^2)(\kappa+\rho c_s^2),\\
b_\pm&=a-c_\pm,\\
c_\pm&=\rho c_s(c_s\pm v_n d),
\end{align}
(i.e., $d\to -d$ for $\Lbf_+ \to \Lbf_-$.)

The above expressions can lead
to cancellation error when quantities are nearly Newtonian, since
then quantities that are close to unity get subtracted from each other.
To avoid this, we rewrite some of the terms.
First, we need the routine that returns the primitive variables
from the conserved ones to
return the quantity $h-1$ accurately,
not simply $h$. This can be done, for example, by using the relation
\begin{equation}
h-1=\frac{\tau+p}{DW}-\frac{v^2 W}{W+1}
\end{equation}
or
\begin{equation}
h-1=\epsilon+\frac{p}{\rho}.
\end{equation}
Then we compute the following ``dangerous'' quantities:
\begin{align}
W-1&=\frac{v^2W^2}{W+1},\\
\label{eq:hmw}
h-W&=(h-1)-(W-1),\\
hW-1&=(h-1)W+(W-1).
\label{eq:hwm1}
\end{align}
Equation \eqref{eq:hwm1} is used in the last term for $\Rbf_3$ and
$\Rbf_\pm$. We rewrite the middle term in $\Lbf_\pm$ as
\begin{multline}
b_\pm-h W(\kappa-\rho c_s^2)(1-v_n^2) = \\
W(1-v_n^2)[\rho c_s^2(W+h)-\kappa(h-W)]-c_\pm.
\end{multline}
We then compute the quantity $(h-W)$ here and in $\Lbf_3$ using
Eq.\ \eqref{eq:hmw}.

Note that $1-v_n^2$ can be computed safely as $1/W^2+v_1^2+v_2^2$, but
this should only be necessary for ultrarelativistic velocities.

\section{Composition Dependence}
\label{sec:composition}

Many current numerical simulations include equation of state and
other effects that depend on the composition of the fluid.
For relativistic applications, we can often assume
nuclear statistical equilibrium. In this case, the equation of state is
described
by just one more thermodynamic variable, usually taken to be the electron
fraction $Y_e$. This quantity satisfies the conservation equation
\begin{equation}
\nabla_a(\rho Y_e u^a)=0.
\label{eq:yecons}
\end{equation}
Using the mass conservation equation \eqref{eq:rhocons}, we can
write this in non-conservative form as
\begin{equation}
u^a \nabla_a Y_e=0.
\end{equation}
So taking as primitive variables $(u^\nu,p,\epsilon,Y_e)$,
we see that Eq.\ \eqref{eq:amatrix} becomes
\begin{equation}
A^a=\begin{bmatrix}
A^a_{(5)} & 0 \\
0 & u^a
\end{bmatrix},
\end{equation}
where $A^a_{(5)}$ is the $5\times 5$ matrix in
Eq.\ \eqref{eq:amatrix}.
Setting up the analog of Eq.\ \eqref{eq:line2}, we see that
there is another degenerate eigenvalue given by $u^a q_a=0$.
The analog of Eq.\ \eqref{eq:degen} now has a zero appended in the $X_6$
position, so the eigenvectors \eqref{eq:eigendegen} are unchanged
(with an extra zero appended corresponding
to the $Y_e$ slot). There is an additional eigenvector corresponding
to taking $X_6=1$:
\begin{equation}
X_4=\begin{bmatrix}
0\\
0\\
0\\
1
\end{bmatrix}.
\label{eq:x4}
\end{equation}

The eigenvectors for the non-degenerate eigenvalues are given by
an equation analogous to Eq.\ \eqref{eq:nondegen}. Showing only
the new terms, we have
\begin{equation}
\begin{bmatrix}
\ddots & & & 0\\
& \ddots & & 0\\
& & \ddots & 0\\
\vphantom{\ddots}0 & 0 & 0 & u^a q_a
\end{bmatrix}
\begin{bmatrix}
\vphantom{\ddots} X^\nu\\
\vphantom{\ddots} X_4\\
\vphantom{\ddots} X_5\\
\vphantom{\ddots} X_6
\end{bmatrix}=0.
\end{equation}
Thus $X_6=0$ and the eigenvectors $X_\pm$ are unchanged (with
a zero appended).

The transformation to the Eulerian frame only affects the middle (vector)
element of each eigenvector, so these eigenvectors will be essentially
the same as before.

\subsection{Transformation to $(\rho,\epsilon,Y_e)$}

Equations \eqref{eq:kappa} and \eqref{eq:inversekappa} become
\begin{align}
\label{eq:comp}
\frac{\partial(p,\epsilon,Y_e)}{\partial(\rho,\epsilon,Y_e)}
&=\begin{bmatrix}
\chi & \kappa &\zeta\\
0 & 1 & 0\\
0 & 0 & 1
\end{bmatrix},\\
\frac{\partial(\rho,\epsilon,Y_e)}{\partial(p,\epsilon,Y_e)}
&=\begin{bmatrix}
1/\chi & -\kappa/\chi & -\zeta/\chi\\
0 & 1 & 0\\
0 & 0 & 1
\end{bmatrix}.
\label{eq:inversecomp}
\end{align}
Here we have defined
\begin{equation}
\zeta=\left.\frac{\partial p}{\partial Y_e}\right|_{\rho,\epsilon}.
\end{equation}
Applying the transformation \eqref{eq:inversecomp} to the bottom 3
components of Eq.\ \eqref{eq:x4} (which is also the Eulerian eigenvector)
gives the eigenvector corresponding
to the new primitive variables:
\begin{equation}
\xbf_4=\begin{bmatrix}
0\\
-\zeta\\
0\\
\chi
\end{bmatrix}.
\label{eq:x4new}
\end{equation}
We can alternatively make a linear combination of this eigenvector
with the degenerate eigenvector $\xbf_3=[0,\,-\kappa,\,\chi,\,0]^T$
of Eq.\ \eqref{eq:fonteigen} and use instead
\begin{equation}
\xbf_4=\begin{bmatrix}
0\\
0\\
-\zeta\\
\kappa
\end{bmatrix}.
\label{eq:x4alt}
\end{equation}
The left eigenvector corresponding to Eq.\ \eqref{eq:x4new} or
\eqref{eq:x4alt} is
\begin{equation}
\lbf_4=\begin{bmatrix}
0 & 0 & 0 & 1/\chi
\end{bmatrix}
\quad\text{or}\quad
\lbf_4=\begin{bmatrix}
0 & 0 & 0 & 1/\kappa
\end{bmatrix}.
\end{equation}
The other left eigenvectors are the same as those given in
Eqs.\ \eqref{eq:leftfontorig} and \eqref{eq:altleft} with an extra term
appended. The simplest way to find the modifications is
to require the left eigenvectors to be orthonormal to the right
eigenvectors.
The extra terms are shown shaded below:

\begin{widetext}
\begin{equation}
\begin{split}
\lbf_{1,2}&=\frac{1}{1-v_n^2}\begin{bmatrix}
(1-v_n^2)t_i^{(1,2)}+v_n v_{1,2} s_i
&
\dfrac{\chi v_{1,2}}{\rho h W^2}
&
\dfrac{\kappa v_{1,2}}{\rho h W^2}
&
\colorbox{mygray}{$\dfrac{\zeta v_{1,2}}{\rho h W^2}$}
\end{bmatrix},\\
\lbf_3&=\frac{1}{h c_s^2}\begin{bmatrix}
0 &
-\dfrac{p}{\rho^2} & 1
 & \colorbox{mygray}{$\dfrac{\zeta}{\kappa}$}
\end{bmatrix},\\
\lbf_\pm&=\pm\frac{1}{(1-v_n^2)(y_- - y_+)}
\begin{bmatrix}
(y_\mp-v_n)s_i &
-\dfrac{(1-y_\mp v_n)\chi}{\rho h W^2} &
-\dfrac{(1-y_\mp v_n)\kappa}{\rho h W^2}
& \colorbox{mygray}{$\dfrac{\zeta}{\rho h W^2 c_s}$}
\end{bmatrix}.
\end{split}
\end{equation}
\end{widetext}

\subsection{Transformation to Conserved Variables}

We see from Eq.\ \eqref{eq:yecons} that the conserved quantity corresponding
to $Y_e$ is $D Y_e$.
To transform from the primitive variables $(u^a,\rho,\epsilon,Y_e)$ to
the conserved variables $(S^a,D,\tau,D Y_e)$, the first step
is to transform from $u^a$ to $v^a$ as in \S\ref{sec:eulerian}.
As before, this only affects the definition of the tangential
vectors $t^b$ and the vector parts of
$\xbf_\pm$, which take on the form
\eqref{eq:xpmv}. Next we apply the transformation matrix whose entries were
computed in Eq.\ \eqref{eq:transtocons}. We also need the additional
entries
\begin{equation}
\begin{gathered}
\frac{\partial S^a}{\partial Y_e}=\zeta W^2 v^a,\quad
\frac{\partial D}{\partial Y_e} = 0,\quad
\frac{\partial \tau}{\partial Y_e}=\zeta(W^2-1),\\
\frac{\partial(D Y_e)}{\partial v^b}=\rho W^3 Y_e v_b,\quad
\frac{\partial(D Y_e)}{\partial \rho}=W Y_e,\quad\\
\frac{\partial(D Y_e)}{\partial \epsilon}=0,\quad
\frac{\partial(D Y_e)}{\partial Y_e}=\rho W.
\end{gathered}
\label{eq:additional}
\end{equation}

The eigenvectors $\Rbf_{1,2}$, $\Rbf_{3}$, and $\Rbf_\pm$
of Eq.\ \eqref{eq:consright} are unchanged
after the transformation, except for each having an additional element appended.
These additional elements are shaded below:

\begin{widetext}
\begin{equation}
\Rbf_{1,2}=
\begin{bmatrix}
h(t^{(1,2)}_i+2W^2v_{1,2}v_i)\\[5pt]
W v_{1,2}\\[5pt]
W(2hW-1)v_{1,2}\\[5pt]
\colorbox{mygray}{$W v_{(1,2)}Y_e$}
\end{bmatrix},
\quad
\Rbf_{3}=
\begin{bmatrix}
hW(\kappa-\rho c_s^2)v_i\\[5pt]
\kappa\\[5pt]
hW(\kappa-\rho c_s^2)-\kappa\\[5pt]
\colorbox{mygray}{$\kappa Y_e$}
\end{bmatrix},
\quad
\Rbf_\pm=
\begin{bmatrix}
hW\left(v_i \pm\dfrac{c_s}{d}s_i\right)\\[5pt]
1\\[3pt]
h W\left(1\pm\dfrac{c_sv_n}{d}\right) -1\\[5pt]
\colorbox{mygray}{$Y_e$}
\end{bmatrix}.
\label{eq:consrightye}
\end{equation}
\end{widetext}

The eigenvector $X_4$ (Eq.\ \ref{eq:x4alt}) is transformed to
\begin{equation}
\Rbf_4=\begin{bmatrix}
-\zeta\rho W^2v^a\\
0\\
-\zeta\rho W^2\\
\kappa \rho W
\end{bmatrix}
\to
\begin{bmatrix}
v^a\\
0\\
1\\
-\dfrac{\kappa}{\zeta W}
\end{bmatrix},
\label{eq:x4first}
\end{equation}
where we have renormalized the eigenvector to get the second expression.

The left eigenvectors are computed using the inverse transformation
matrix, which is
computed in Appendix \ref{app:inverse_ye}. Using the transformation
\eqref{eq:invertible}, we get a fairly simple result.
The additional terms now appearing in 
Eq. \eqref{eq:left_cons} are shown shaded below:
\begin{equation}
\begin{split}
\Lbf_{1,2}&=\frac{1}{h (1-v_n^2)}
\left[
\begin{matrix}
v_{1,2}\,v_n s_i+(1-v_n^2) t^{(1,2)}_i\\
-v_{1,2}\\
-v_{1,2}\\
\colorbox{mygray}{$0$}
\end{matrix}
\right]^T,\\
\Lbf_{3}&=\frac{1}{\rho h c_s^2}
\left[
\begin{matrix}
W v_i \\[1pt]
h-W \colorbox{mygray}{${}+\zeta Y_e/\kappa$}\\
-W\\
\colorbox{mygray}{$-\zeta /\kappa$}
\end{matrix}
\right]^T,\\
\Lbf_{\pm}&=\frac{1}{2\rho h W c_s^2 (1-v_n^2)}\\
&\quad
\left[
\begin{matrix}
-a v_i+ \rho c_s(c_s v_n\pm d) s_i\\[2pt]
b_\pm-h W(\kappa-\rho c_s^2 \colorbox{mygray}{${}+\zeta Y_e/h$})(1-v_n^2)\\
b_\pm\\
\colorbox{mygray}{$\zeta W (1-v_n^2)$}
\end{matrix}
\right]^T.
\end{split}
\label{eq:left_ye}
\end{equation}
Finally, the additional degenerate eigenvector is
\begin{equation}
\Lbf_4=\frac{\zeta W}{\kappa}\begin{bmatrix}
0 & Y_e & 0 & -1
\end{bmatrix}.
\end{equation}

\section{Discussion}
\label{sec:discussion}

The characteristic decomposition for GRMHD is not known in a form
useful for current numerical simulations. In this paper,
we have presented a concise method to derive the decomposition
for relativistic hydrodynamics in conservation form. In Paper II,
we show that this method works for GRMHD as well.

The method presented here relies on
a new transformation technique called a
quasi-invertible transformation.
We recover the known results for relativistic
hydrodynamics in somewhat simpler form
and without the need for computer algebra.

In \S\S\ref{sec:changing} -- \ref{sec:lefttrans}, we discuss
the transformation of eigenvectors under various circumstances.
In particular, we introduce quasi-invertible transformations
with the simple example of going from 4-velocity to a spatial 3-velocity.
The rectangular matrix of the transformation does not have a proper inverse.
Nevertheless, it does have a one-sided inverse that is not unique.
A simple prescription picks out a convenient unique choice. We then
apply transformations to get the eigenvectors in terms of the conserved
variables used in simulations. These eigenvectors are listed in
\S\ref{sec:conservative}.

A new result in this paper is the characteristic decomposition
when the the evolution tracks the composition
of the fluid in nuclear statistical equilibrium.
These results are obtained with simple
extensions of the results where composition changes are ignored
and are presented in \S\ref{sec:composition}.

The immediate next step is to apply the methods presented here
to GRMHD. This is carried out in Paper II~\cite{teukolsky2025b}.
The technical challenge there will be to develop the quasi-invertible
transformation that takes the 4-d comoving magnetic field to the
3-d spatial Eulerian magnetic field. This turns out to be
non-trivial. The final result of that paper is
the complete characteristic decomposition for GRMHD in terms
of the conserved variables used in numerical simulations.
These results mean that
MHD systems can now be handled with the most accurate known
computational methods, including full-wave Riemann solvers
like Roe or Marquina solvers.
We no longer have to rely on crude approximate treatments of
numerical fluxes and boundary conditions in GRMHD.

\begin{acknowledgments}
We have greatly benefited from the Mathematica notebooks that
accompany Refs.~\cite{schoepe2018} and \cite{hilditch2019}.
While the details of our calculations are very different,
the Mathematica manipulations have many similarities.
These notebooks use the xTensor package~\cite{xtensor}.
This work was supported in part by NSF grants
PHY-2308615 and OAC-2513338 and by NASA award 80NSSC26K0340 at Cornell.
This work was also supported in part by the Sherman
Fairchild Foundation at Caltech and Cornell.
\end{acknowledgments}

\appendix

\section{Speed of Sound}
\label{app:a}

Most equations of state give the pressure as a function of $\rho$
and $\epsilon$, so we need an expression for the speed of sound
$c_s$ in terms of these variables.
We have
\begin{multline}
c_s^2=\left.\frac{\partial p}{\partial e}\right|_s
=\frac{\partial(p,s)}{\partial(e,s)}
=\frac{\partial(p,s)}{\partial(\rho,\epsilon)}
\frac{\partial(\rho,\epsilon)}{\partial(e,s)}\\
=
\begin{vmatrix}
\dfrac{\partial p}{\partial \rho}
&
\dfrac{\partial p}{\partial \epsilon}
\\[10pt]
\dfrac{\partial s}{\partial \rho}
&
\dfrac{\partial s}{\partial \epsilon}
\end{vmatrix}
\begin{vmatrix}
\dfrac{\partial e}{\partial \rho}
&
\dfrac{\partial e}{\partial \epsilon}
\\[10pt]
\dfrac{\partial s}{\partial \rho}
&
\dfrac{\partial s}{\partial \epsilon}
\end{vmatrix}^{-1}.
\label{eq:determinant}
\end{multline}
Get the derivatives of $e$ and $s$ from Eqs.\ \eqref{eq:e} and \eqref{eq:ds}.
Thus
\begin{equation}
c_s^2=
\begin{vmatrix}
\chi & \kappa\\
-\dfrac{p}{\rho^2 T} & \dfrac{1}{T}
\end{vmatrix}
\begin{vmatrix}
1+\epsilon & \rho\\
-\dfrac{p}{\rho^2 T} & \dfrac{1}{T}
\end{vmatrix}^{-1}
=
\frac{1}{h}\left(\chi+\kappa\frac{p}{\rho^2}\right).
\label{eq:cs2}
\end{equation}

A more compact derivation of this expression is possible (e.g.,
Ref.~\cite{rezzollabook}):
\begin{multline}
c_s^2=\left.\frac{\partial p}{\partial e}\right|_s
=\left.\frac{\partial \rho}{\partial e}\right|_s
\left.\frac{\partial p}{\partial \rho}\right|_s
=\frac{1}{h}\left.\frac{\partial p}{\partial \rho}\right|_s\\
=\frac{1}{h}\left[
\left.\frac{\partial p}{\partial \rho}\right|_\epsilon
+
\left.\frac{\partial \epsilon}{\partial \rho}\right|_s
\left.\frac{\partial p}{\partial \epsilon}\right|_\rho
\right].
\end{multline}
We get the derivatives
\begin{equation}
\left.\frac{\partial \rho}{\partial e}\right|_s=\frac{1}{h},
\qquad
\left.\frac{\partial \epsilon}{\partial \rho}\right|_s=\frac{p}{\rho^2},
\end{equation}
from the First Law in the forms $de=h\,d\rho+\rho T\,ds$ and
Eq.\ \eqref{eq:ds}, and hence recover Eq.\ \eqref{eq:cs2}. However,
we prefer the Jacobian method used above because it is general
and directly introduces the required independent variables.

\section{Quasi-invertible Transformation for GRMHD}
\label{app:quasiinv}
For GRMHD, to transform from the comoving frame to the Eulerian frame,
we need the transformation from  $\{u^a,b^a\}$ to $\{v^a,B^a\}$, as well as the
inverse transformation. Here $b^a$ is the comoving magnetic field while
$B^a$ is the Eulerian magnetic field.
The second equation in
\eqref{eq:simpleqi} can be regarded as the definition of a left
inverse $\partial U/\partial U_{\rm old}$
for $\partial U_{\rm old}/\partial U$.
Since the relation between $u^a$
and $v^a$ does not involve the magnetic field, this equation is explicitly
\begin{align}
\frac{\partial(v^a,B^a)}{\partial (u^c,b^c)}
\frac{\partial(u^c,b^c)}{\partial (v^b,B^b)}&=
\begin{bmatrix}
\dfrac{\partial v^a}{\partial u^c} & 0\\[7pt]
\dfrac{\partial B^a}{\partial u^c} &
\dfrac{\partial B^a}{\partial b^c}
\end{bmatrix}
\begin{bmatrix}
\dfrac{\partial u^c}{\partial v^b} & 0\\[7pt]
\dfrac{\partial b^c}{\partial v^b} & 
\dfrac{\partial b^c}{\partial B^b}
\end{bmatrix}\notag\\[2pt]
&=\begin{bmatrix}
\gamma^a{}_b & 0\\
0 & \gamma^a{}_b
\end{bmatrix}.
\label{eq:jacobian}
\end{align}
This gives two conditions on the magnetic field part of the
transformation:
\begin{align}
\label{eq:first}
\dfrac{\partial B^a}{\partial u^c}\dfrac{\partial u^c}{\partial v^b}
+\dfrac{\partial B^a}{\partial b^c}\dfrac{\partial b^c}{\partial B^b}
&=0,\\
\dfrac{\partial B^a}{\partial b^c}\dfrac{\partial b^c}{\partial B^b}
&=\gamma^a{}_b.
\label{eq:second}
\end{align}
We show in Paper II how to satisfy these equations.

The first equation in \eqref{eq:simpleqi} multiplies the matrices in
Eq.\ \eqref{eq:jacobian} in the opposite order as a convenient
way to pick out a unique inverse, giving the two conditions
\begin{align}
\label{eq:invfirst}
\dfrac{\partial b^a}{\partial v^c}
\dfrac{\partial v^c}{\partial u^b}
+
\dfrac{\partial b^a}{\partial B^c}
\dfrac{\partial B^c}{\partial u^b}
&\stackrel{?}{=}0,\\
\dfrac{\partial b^a}{\partial B^c}
\dfrac{\partial B^c}{\partial b^b}
&=h^a{}_b.
\label{eq:invsecond}
\end{align}
The question mark over the equals sign in Eq.\ \eqref{eq:invfirst} is there
because we will find that we cannot in fact accomplish this requirement.
Fortunately, we can find an alternative requirement that is equally
effective.
It turns out that we can allow the
right-hand side of \eqref{eq:invfirst} to be proportional to $u^a$.
The resulting expression for ${\partial U_{\rm old}}/{\partial U}$
then contains a term proportional to $u^a$ that does not contribute
when multiplied on the left by $\left(A^a_{\rm old}n_a\right)$ in
Eq.\ \eqref{eq:ltrans}, so the transformation is quasi-invertible.

\section{Barotropic Equation of State}
\label{app:c}

We show here that for a barotropic equation of state, i.e., when
$p=p(e)$, then $\kappa-\rho c_s^2=0$. First, we repeat the proof
of~\cite{ibanez2012}. Taking $(\rho,\epsilon)$ as the independent variables,
we have
\begin{align}
\kappa-\rho c_s^2&=\kappa -\frac{\rho}{h}\left(
\chi+\frac{p}{\rho^2}\kappa\right)\notag\\
&=\frac{1}{h}\left[\kappa\left(h-\frac{p}{\rho}\right)-\rho\chi\right]\notag\\
&=\frac{1}{h}\left(\frac{e}{\rho}\kappa-\rho\chi\right)\label{eq:kaprho}\\
&=\frac{1}{h}\left(\frac{\partial e}{\partial\rho}\frac{\partial p}{\partial
\epsilon}-\frac{\partial e}{\partial\epsilon}\frac{\partial p}{\partial \rho}
\right).
\end{align}
Here we have introduced the derivatives of $e$ from the expression
$e=\rho(1+\epsilon)$.
Thus
\begin{equation}
\kappa-\rho c_s^2=0\quad\Leftrightarrow\quad
\frac{\partial(p,e)}{\partial(\rho,\epsilon)}=0\quad\Leftrightarrow\quad
p=p(e).
\end{equation}

Next we give a somewhat more direct proof. A barotropic equation of state
has constant entropy, so consider changing variables from $(e,s)$ to
$(\rho,\epsilon)$. Then for a general equation of state, following
Eqs.~\eqref{eq:determinant} and \eqref{eq:cs2} we have
\begin{multline}
\left.\frac{\partial p}{\partial s}\right|_e
=\frac{\partial(p,e)}{\partial(s,e)}
=\frac{\partial(p,e)}{\partial(\rho,\epsilon)}
\frac{\partial(\rho,\epsilon)}{\partial(s,e)}\\
=
\begin{vmatrix}
\dfrac{\partial p}{\partial \rho}
&
\dfrac{\partial p}{\partial \epsilon}
\\[10pt]
\dfrac{\partial e}{\partial \rho}
&
\dfrac{\partial e}{\partial \epsilon}
\end{vmatrix}
\begin{vmatrix}
\dfrac{\partial s}{\partial \rho}
&
\dfrac{\partial s}{\partial \epsilon}
\\[10pt]
\dfrac{\partial e}{\partial \rho}
&
\dfrac{\partial e}{\partial \epsilon}
\end{vmatrix}^{-1}
=
\begin{vmatrix}
\chi & \kappa\\
1+\epsilon & \rho
\end{vmatrix}
\begin{vmatrix}
-\dfrac{p}{\rho^2 T} & \dfrac{1}{T}\\
1+\epsilon & \rho
\end{vmatrix}^{-1}\\
=\frac{\rho\chi-\kappa(1+\epsilon)}{-p/\rho T-(1+\epsilon)/T}
=\frac{T}{h}\left(\frac{e}{\rho}\kappa-\rho\chi\right)
=T(\kappa-\rho c_s^2),
\end{multline}
where we have used Eq.~\eqref{eq:kaprho} to get the last equality.
Thus if a general equation of state $p=p(e,s)$ has constant entropy, so
that it reduces to $p=p(e)$, then $\kappa-\rho c_s^2=0$.

\section{Inverse Transformation for Conserved Variables}
\label{app:inverse}
The inverse transformation matrix \eqref{eq:inversecons} is explicitly
\begin{widetext}
\begin{equation}
\frac{\partial(v^b,\rho,\epsilon)}{\partial(S^a,D,\tau)}=
\frac{1}{\rho h f_1}
\begin{bmatrix}
\dfrac{f_1}{W^2}\delta_a{}^b-\dfrac{\kappa+\rho c_s^2}{\rho}v_av^b &
\dfrac{W(\kappa+\rho)-h(\kappa-\rho c_s^2)}{\rho W}v_a &
\dfrac{\kappa+\rho}{\rho}v_a\\
[(W^2-1)\kappa+W^2\rho]v^b & \dfrac{(h-W)(W^2-1)\kappa}{W}-\rho f_3 &
-(W^2-1)(\kappa+\rho)\\
\dfrac{pW^2+\rho f_2}{\rho}v^b &
-\dfrac{p}{\rho}f_3+\dfrac{(h-W)f_2}{W} &
-\dfrac{p}{\rho}(W^2-1)-f_2
\end{bmatrix},
\label{eq:inverseconsmatrix}
\end{equation}
\end{widetext}
where
\begin{align}
f_1&=-[W^2+c_s^2(1-W^2)],\\
f_2&=hW^2-\chi(W^2-1),\\
f_3&=hW-1+W^2.
\end{align}

\section{Inverse Transformation for Conserved Variables with Composition
Dependence}
\label{app:inverse_ye}
Since we already have the transformation \eqref{eq:transtocons}
and its inverse \eqref{eq:inverseconsmatrix},
the simplest way to compute the inverse transformation matrix
$\partial(v^b,\rho,\epsilon,Y_e)/\partial(S^a,D,\tau,D Y_e)$ is
to use the formulas for the inverse of a partitioned matrix.
Suppose a matrix $A$ and its inverse $A^{-1}$ are partitioned as
\begin{equation}
A=
\begin{bmatrix}
P & Q\\
R & S
\end{bmatrix},
\quad
A^{-1}=\begin{bmatrix}
\widetilde P & \widetilde Q\\
\widetilde R & \widetilde S
\end{bmatrix},
\end{equation}
with $P$ and $S$ square (but not necessarily the same size).
Then we can compute the sub-matrices of the inverse by computing in
sequence
\begin{align}
\label{eq:stilde}
\widetilde S&=(S-R\cdot P^{-1}\cdot Q)^{-1},\\
\label{eq:rtilde}
\widetilde R&=-\widetilde S\cdot (R\cdot P^{-1}),\\
\label{eq:qtilde}
\widetilde Q & = -(P^{-1}\cdot Q)\cdot \widetilde S,\\
\label{eq:ptilde}
\widetilde P &= P^{-1}-(P^{-1}\cdot Q)\cdot \widetilde R.
\end{align}
(There is a similar formula that starts with computing $\widetilde P$ from
$S^{-1}$.) We set
\begin{equation}
\begin{split}
P&= \text{Eq.\ \eqref{eq:transtocons}},\\
Q&=\begin{bmatrix}
0\\
\zeta W^2 v_b\\
\zeta(W^2-1)
\end{bmatrix},\\
R&=\begin{bmatrix}
W Y_e & W^3 Y_e v^a & 0
\end{bmatrix},\\
S&= \rho W.
\end{split}
\end{equation}
Here $Q$, $R$, and $S$ come from Eq.\ \eqref{eq:additional}.

Using Eqs.\ \eqref{eq:stilde} -- \eqref{eq:ptilde},
we find that we can represent the inverse by adding the following matrix to 
Eq.\ \eqref{eq:inverseconsmatrix} padded with an extra row and column
of zeros:
\begin{equation}
\frac{1}{\rho h f_1}
\begin{bmatrix}
0 & -\dfrac{\zeta Y_ev_a}{\rho W} & 0& \dfrac{\zeta v_a}{\rho W}\\[9pt]
0 & \dfrac{\zeta Y_e(W^2-1)}{W} & 0 & -\dfrac{\zeta (W^2-1)}{W}\\[9pt]
0 & \dfrac{\zeta Y_e p(W^2-1)}{W} & 0 &  -\dfrac{\zeta Y_e p(W^2-1)}{\rho^2 W}\\[9pt]
0 & -\dfrac{Y_e h f_1}{W} & 0 & \dfrac{h f_1}{W}
\end{bmatrix}.
\end{equation}

\newpage  

\bibliography{references}

\end{document}